\documentclass[aps,prl,twocolumn,notitlepage]{revtex4-1}
\usepackage[a4paper,top=1.75cm,bottom=1.75cm,left=1.75cm,right=1.75cm]{geometry}
\usepackage{amsmath}
\usepackage{amssymb}
\usepackage{xfrac}
\usepackage{bm}
\usepackage{enumerate}
\usepackage{hyperref}
\usepackage{cleveref}[2012/02/15]% v0.18.4; 
\crefformat{footnote}{#2\footnotemark[#1]#3}
\usepackage{mathptmx}
\usepackage[utf8]{inputenc}
\usepackage{float}
\usepackage{graphicx}
\DeclareGraphicsExtensions{.eps, .pdf, .jpg, .png}
\usepackage[usenames,dvipsnames]{color}
\graphicspath{{./}}
\def\kb{k_{\text{B}}}
\newcommand{\betawc}[1]{\beta_{#1}^{\text{wc}}}
\newcommand{\betasc}[1]{\beta_{#1}^{\text{sc}}}

\newcommand{\scb}[1]{\beta^{\text{sc}}_{#1}}

\newcommand{\mbar}{\,\mbox{bar}}

\newcommand{\pPCP}{p_{\mbox{\tiny PCP}}}
\newcommand{\TAB}{T_{\mbox{\tiny AB}}}
\newcommand{\vOmega}{\pmb{\Omega}}
\newcommand{\vell}{\mbox{\boldmath$\ell$}}
\def\ns{\negthickspace}

\newcommand*\rfrac[2]{{}^{#1}\!/_{#2}}
\def\pder#1#2{\mbox{$\displaystyle\frac{\partial #1}{\partial #2}$}}
\def\cO{{\mathcal O}}
\newcommand{\He}{$^3$He}
\newcommand{\Heb}{$^3$He-B}
\newcommand{\Hea}{$^3$He-A}
\newcommand{\Hefour}{$^4$He}
\def\vH{{\bf H}}
\def\vM{{\bf M}}
\def\vp{{\bf p}}
\newcommand{\grad}{\mbox{\boldmath$\nabla$}}
\def\cF{{\mathcal F}}
\def\Tr#1{\mbox{Tr}\big\{#1\big\}}
\newcommand{\be}{\begin{equation}}
\newcommand{\ee}{\end{equation}}
\newcommand{\ber}{\begin{eqnarray}}
\newcommand{\eer}{\end{eqnarray}}
\def\nicefrac#1#2{\genfrac{}{}{}{1}{#1}{#2}}
\newcommand{\onethird}{\frac{\mbox{\small 1}}{\mbox{\small 3}}}
\def\va{{\bf a}}
\def\vj{{\bf j}}
\def\vm{{\bf m}}
\def\vn{{\bf n}}
\def\vr{{\bf r}}
\def\vs{{\bf s}}
\def\vu{{\bf u}}
\def\vv{{\bf v}}
\def\vx{{\bf x}}
\def\vy{{\bf y}}
\def\vz{{\bf z}}
\def\vH{{\bf H}}
\def\vJ{{\bf J}}
\def\vL{{\bf L}}
\def\vS{{\bf S}}
\def\cA{{\mathcal A}}
\def\cD{{\mathcal D}}
\def\cF{{\mathcal F}}
\def\cL{{\mathcal L}}
\def\point#1#2{{\tt #1}_{\mbox{\tiny #2}}}
\def\orbital{{\tt SO(3)_{\text{L}}}}
\def\spin{{\tt SO(3)_{\text{S}}}}
\def\gauge{{\tt U(1)_{\text{N}}}} 
 
\def\time{{\tt T}}
\def\spinorbital{{\tt SO(3)_{\text{L+S}}}}
\begin{document}
\title{The Vortex Phase Diagram of Rotating Superfluid $^3$He-B}
\author{Robert C. Regan}
\email{robertregan2018@u.northwestern.edu}
\author{J. J. Wiman}
\altaffiliation{MC2, Chalmers University, Gothenburg, Sweden}
\email{jwiman@chalmers.se}
\author{J. A. Sauls}
\email{sauls@northwestern.edu}
\affiliation{Department of Physics, Northwestern University, Evanston, IL 60208}
\date{\today}
%----------------------------------------------------------------------------------------
%
\begin{abstract}
We present the first theoretical calculation of the pressure-temperature-field phase diagram for the vortex phases of rotating superfluid \Heb.
Based on a strong-coupling Ginzburg-Landau functional that accounts for the relative stability of the bulk A and B phases of \He\ at all pressures, we report calculations for the internal structure and free energies of distinct broken-symmetry vortices in rotating superfluid \Heb.
Theoretical results for the equilibrium vortex phase diagram in zero field and an external field of $H=284\,\mbox{G}$ parallel to the rotation axis, $\vH\parallel\pmb{\Omega}$, are reported, as well as the supercooling transition line, $T^{*}_{\text{V}}(p,H)$.
In zero field the vortex phases of \Heb\ are separated by a first-order phase transition line $T_{\text{V}}(p)$ that terminates on the bulk critical line $T_{c}(p)$ at a triple point.
The low-pressure, low-temperature phase is characterized by an array of singly-quantized vortices that spontaneously breaks axial rotation symmetry, exhibits anisotropic vortex currents and an axial current anomaly (D-core phase).
The high-pressure, high-temperature phase is characterized by vortices with both bulk A phase and $\beta$ phase in their cores (A-core phase). We show that this phase is metastable and supercools down to a minimum temperature, $T^{*}_{\text{V}}(p,H)$, below which it is globally unstable to an array of D-core vortices.
For $H\gtrsim 60\,\mbox{G}$ external magnetic fields aligned along the axis of rotation increase the region of stability of the A-core phase of rotating \Heb, opening a window of stability down to low pressures.
These results are compared with the experimentally reported phase transitions in rotating \Heb.
\end{abstract}
\maketitle
%---------------------------------------------------------------------------------------

\section{Introduction} \label{sec:introduction}
%\noindent{\it Introduction} --
%
The velocity field of a superfluid is irrotational. Nevertheless superfluids can approximate solid body rotation when confined in a container rotating at constant angular speed. Co-rotation is achieved by the nucleation of an array of vortices, each of which possesses a quantum of circulation.
In superfluid \Hefour, or in a spinless, s-wave BCS superfluid the condensate wavefunction, or \emph{order parameter}, is a complex scalar field. The quantum of circulation is then $\kappa=h/M$, where $h$ is Planck's constant and $M$ is the mass of the fundamental constituent of the condensate~\cite{ons49,fey55}.

Quantization of circulation reflects the single-valuedness of the condensate wave function, and non-trivial topology of the degeneracy space of the order parameter manifold.
In a cylindrical container vortices align parallel to the angular velocity, $\vOmega$, and co-rotation is achieved at an average areal vortex density of $n_{\text{V}}=2\Omega/\kappa$~\cite{fey55}.
Long-range, repulsive interactions lead to a two-dimensional lattice of rectilinear vortices, which for axially symmetric vortices is a two-dimensional hexagonal lattice with inter-vortex spacing, $d$, determined by $d^{2}=\kappa/\sqrt{3}\Omega$, which depends only on fundamental constants and the speed of rotation.
Thus, for \Hefour, or an isotropic BCS superfluid, once a sufficient number of axially symmetric vortices nucleate to form the vortex lattice no further symmetry breaking phase transition is expected until the density approaches a critical density at which neighboring vortex cores overlap and superfluidity is destroyed at an upper critical rotation speed of $\Omega_{c_2}\approx\kappa/\xi^2$. 
For superfluid \He\, which is a BCS condensate of Cooper pairs with $\kappa=h/2m_3\approx 0.066\,\mbox{mm}^2/s$~\cite{zie93} and a core size $\xi\approx 20-80\,\mbox{nm}$ over the presssure range $p=0-34\,\mbox{bar}$, $\Omega_{c_2}\gtrsim 10^{7}\,\mbox{s}^{-1}$, which is experimentally inaccessible.

However, the ground state of superfluid \He\ is a time-reversal invariant, spin-triplet, p-wave topological superfluid that breaks orbital and spin rotation symmetries, $\orbital\times\spin$, in addition to $\gauge$ gauge symmetry, but is invariant under joint spin and orbital rotations, $\spinorbital$~\cite{vollhardt90}.
The resulting degeneracy space allows for a number of unique topologically stable defects~\cite{sal87,lou99}, including quantized vortices with different internal core structures~\cite{sal83,thu86}.
This opens the possibility of multiple superfluid phases characterized by distinct vortex structures. 

Indeed experimental evidence of multiple vortex phases in rotating \Heb\ was reported soon after the first rotating milli-Kelvin cryostat in Helsinki was operational~\cite{ikk82,hak83}.
Using nuclear magnetic resonance (NMR) spectroscopy the vortex array in rotating superfluid \Heb\ was detected as a change in the level spacing of the spin-wave bound-state spectrum proportional to the vortex density, $\Delta\omega_{\text{sw}}\propto n_{\text{V}}\propto\Omega$, for rotation speeds, $\Omega = 0.2-1.7\,\mbox{rad/s}$~\cite{hak83,hak83a}.
A discontinuity in $\Delta\omega_{\text{sw}}/\Omega$ at $T_{\text{V}}^*\approx 0.6\,T_c$ was the signature of a first-order phase transition associated with the vortex array~\cite{ikk82,hak83}. 
The rotation-induced NMR bound-state frequency shift also depends on the \emph{relative} orientation of the NMR field and the angular velocity, i.e. there is a gyromagnetic splitting, $\delta\omega_{\text{gyro}}\propto n_{\text{V}}\vH\cdot\vM_{\text{V}}$, indicative of an intrinsic magnetization generated by the circulation of the spin-triplet Cooper pairs in the region of the vortex-core, $\vM_{\text{V}}=M_{\text{V}}\,\hat{\pmb{\Omega}}$, the magnitude of which depends on the internal structure of the vortex core~\cite{hak83}.

There are two equilibrium phases of \Heb\ under rotation. Over most of the p-T phase diagram rotating \Heb\ is believed to be defined by an array of line defects that are singly quantized mass vortices, each of which spontaneously breaks rotational symmetry, manifest by an anisotropic, double-core structure (D-core) of the Cooper pair density. This structure for the low-temperature, lower pressure vortex phase was discovered by Thuneberg based on numerical solutions of the GL equations that did not constrain the order parameter to be axially symmetric~\cite{thu86}.
At higher temperatures and pressures the phase of rotating \Heb\ is believed to be an array of vortices in which local rotational symmetry is restored, but time-reversal symmetry is broken via the nucleation of both the chiral A phase and the non-unitary $\beta$ phase in the core. 
The stability of \Heb\ with an array of A-core vortices with ferromagnetic cores was argued based on a symmetry classification of axially symmetric B phase vortices and the observation of a measureable gyromagnetic effect from vortices in rotating \Heb\ by Salomaa and Volovik~\cite{sal83}.
However, a quantitative theory of the relative stability of the A-core and D-core vortex phases as a function of pressure, temperature and magnetic field was beyond the scope of existing theory of superfluid \He\ until now.

%--------------------------------------------------------------------------------------
\begin{figure}[t] 
\centering
\includegraphics[width=\columnwidth]{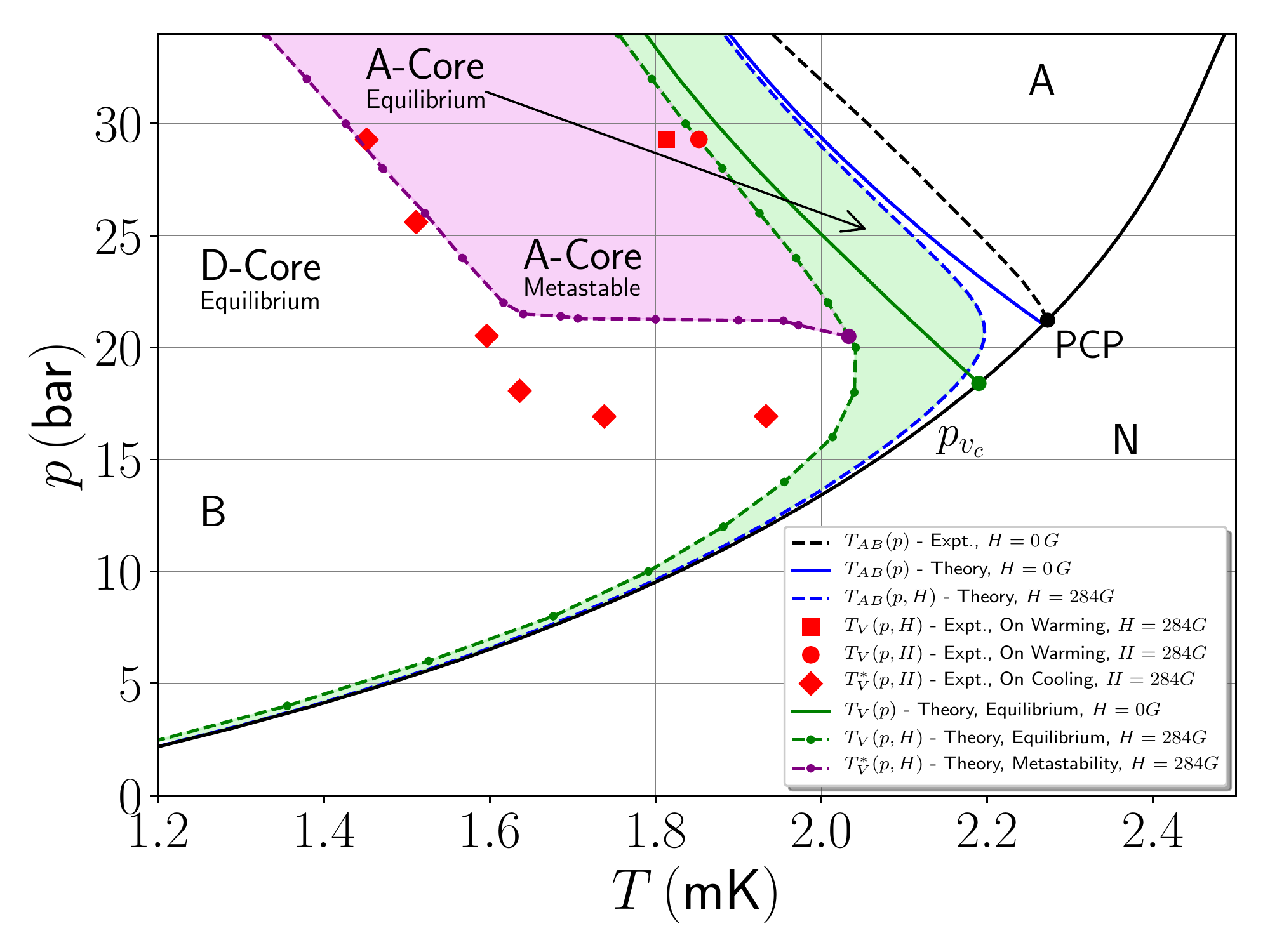}
\caption{The vortex core transition line, $T_{\text{V}}(p,H)$, for $H=0$ G (solid green) separating the A-core and D-core vortex phases of \Heb\ terminates at a triple point $(p_{v_c},T_{v_c})=(18.40\,\mbox{bar},2.19\,\mbox{mK})$. In a magnetic field the vortex core transition line extends to low temperatures down to $p=0$ bar, as shown for $H=284$ G (dashed green). For comparison the Bulk AB transition lines for are shown in blue for $H=0$ G (solid) and $H=284$ G (dashed). The A-core phase supercools down to the metastability limit, $T_{\text{V}}^{*}(p,H)$, shown as the purple dashed line for $H=284$ G. Experimental data for the transition on cooling (red diamonds) agrees well with the supercooling transition, while the data point at $p=29.3$ bar taken on warming (red square/circle) agrees well with the calculated equilibrium vortex phase transition. The experimental data is from Ref.~\cite{pek84a}.
}
\label{fig-Vortex_Phase_Diagram}
\end{figure}
%--------------------------------------------------------------------------------------

Based on a recent formulation of the strong-coupling Ginzburg-Landau theory that accounts for the relative stability of the bulk A and B phases of \He\ for all pressures~\cite{wim15,wim19}, we report calculations of the internal structure and energetics of topologically distinct vortices in rotating superfluid \Heb. In particular, we report the first theoretical calculation of the pressure-temperature-field phase diagram for the vortex phases of rotating superfluid \Heb.
Theoretical results for the equilibrium vortex phase diagram in zero-field and in an external field of $H=284\,\mbox{G}$ parallel to the rotation axis, $\vH\parallel\pmb{\Omega}$, are reported, as well as the supercooling transition, $T_{\text{V}}^{*}(p,H)$, defining the region of metastability of the A-core vortex phase.
Central results reported here include the equilibrium phase diagram based on precise numerical solutions of the strong-coupling theory for the vortex phases of rotating \Heb\ shown in Fig.~\ref{fig-Vortex_Phase_Diagram}, as well as the region of a metastable A-core phase. Also shown in Fig.~\ref{fig-Vortex_Phase_Diagram} are the experimental results for the first-order phase transitions between distinct vortex phases in rotating \Heb, both on cooling and on warming. The transitions on cooling for $H=284$ G over a wide pressure range agree with the theoretically determined metastability transition, $T_{\text{V}}^*(p,H)$, at which the A-core phase is globally unstable for pressures $p\gtrsim 20\,\mbox{bar}$. Furthermore, the transition on warming at $p=29.3$ bar and $H=284$ G is in close agreement with our determination of the equilibrium transition line, $T_{\text{V}}(p,H)$, at that pressure and field. We discuss the phase diagram in more detail in Sec.~\ref{sec-Metastable_A-core_Phase}.
These results provide strong theoretical support for the identification of the vortex phases of \Heb\ as those originally proposed: the low-temperature, low-pressure D-core vortex phase by Thuneberg~\cite{thu86}, and the high-pressure, high-temperature phase as an array of A-core vortices. The A-core vortex phase first described by Salomaa and Volovik was originally proposed as the low-temperature vortex phase~\cite{sal83}.

In Sec.~\ref{sec-GL_Theory} we begin with a description of the strong-coupling GL theory that is the basis for our analysis summarized in Fig.~\ref{fig-Vortex_Phase_Diagram}. In Sec.~\ref{sec-Vortex_States} we describe the stationary state vortex solutions of the strong-coupling GL theory, including their topology and broken symmetries. 
We describe the key features of the axi-symmetric A-core vortex phase as well as the non-axi-symmetric D-core vortex, including their internal topology, mass currents and magnetic properties.
Visualization of the amplitude and phase structure of vortex states leads us to identify the \emph{mechanism} responsible for the phase transition to the D-core phase at $T_{\text{V}}(p,H)$. 
In Sec.~\ref{sec-Magnetic_Susceptibility_Vortex_States} we discuss the local magnetic susceptibilities of the A-and D-core vortices, and the resulting field evolution of the equilibrium A-core to D-core transition.
We discuss the metastability of the A-core phase in Sec.~\ref{sec-Metastable_A-core_Phase}, and the analysis underlying the supercooling transition line, $T^{*}_{\text{V}}(p,H)$, shown in Fig.~\ref{fig-Vortex_Phase_Diagram}.
Our numerical results for the stationary states of the free energy functional are based on a fast converging algorithm described in Appendix~\ref{app-l-bfgs}.

%--------------------------------------------------------------------------------------
\vspace*{-3mm}
\section{Ginzburg-Landau Theory}\label{sec-GL_Theory}
\vspace*{-3mm}
\smallskip
The B phase of superfluid \He\ is the p-wave, spin-triplet Balian-Werthamer state that is invariant under joint spin and orbital rotations as well as time reversal, ${\mathsf H}=\spinorbital\times\time$. The corresponding degeneracy space of \Heb\ allows for a unique spectrum of topologically stable defects, including several quantized mass current vortices with distinct broken symmetries~\cite{ohm83,sal83,thu86}. Topological defects often host distinct inhomogeneous phases, confined within their cores, but embedded in the order parameter field of the ground state~\cite{sal85}. Thus, a theoretical description of vortices in rotating \Heb\ requires a theory allowing for all possible realizations of the order parameter for spin-triplet, p-wave pairing.  

The GL theory is formulated as a functional of the order parameter, the amplitude for the condensate of Cooper pairs, $\langle \psi_{\sigma}(\vp)\psi_{\sigma'}(-\vp)\rangle$ in the spin-momentum basis.
For spin-triplet, p-wave Cooper pairs the condensate amplitude can be expressed in terms of a $3\times3$ matrix order parameter, $A_{\alpha i}$, of complex amplitudes that transforms as the vector representation of $\point{SO(3)}{S}$ with respect to the spin index $\alpha=\{x',y',z'\}$, and as the vector representation of $\point{SO(3)}{L}$ with respect to the orbital momentum index $i=\{x,y,z\}$.
The GL free energy functional is expressed in terms of linearly independent invariants constructed from $A_{\alpha i}$, $A_{\alpha i}^{*}$ and their gradients, $\grad_j A_{\alpha i}$ and $\grad_j A_{\alpha i}^{*}$. In particular, the GL functional can be expressed in terms of free energy densities~\cite{thu87,wim15},
\begin{equation}\label{eq-GL_functional}
\cF[A] = \int_{V} d^3r\,
\left(
f_{\mathrm{bulk}}[A] 
+ 
f_{\mathrm{field}}[A]
+ 
f_{\mathrm{grad}}[A] 
\right)
\,,
\end{equation}
where the bulk free energy density is given by one second-order invariant and five fourth-order invariants,
\begin{align} 
f_\mathrm{bulk}[A] 
&=  
\alpha(T) \Tr{A A^{\dagger}}
+\beta_{1} \left|\Tr{A A^{T}}\right|^{2}
\nonumber \\
&
+\beta_{2} \left[\Tr{A A^{\dagger}}\right]^{2}
+\beta_{3}\, \Tr{AA^{T}(AA^{T})^{*}}
\nonumber \\
&
+\beta_{4}\,\Tr{(AA^{\dagger})^{2}}
+\beta_{5}\,\Tr{AA^{\dagger}(AA^{\dagger})^{*}}
\,,
\label{eq-condensation_energy}
\end{align}
where $A^{\dag}$ ($A^{T}$) is the adjoint (transpose) of $A$. 

The nuclear Zeeman energy for spin-triplet pairs also plays a role in the determination of the vortex structure and phase diagram for the vortex phases of rotating \Heb, even for relatively weak fields.
The dominant field-dependent term in the GL functional is a bulk term representing a correction to the nuclear Zeeman energy from the condensate of spin-triplet Cooper pairs,
\begin{equation}\label{eq-Zeeman_energy}
f_\mathrm{field}[A] 
=g_{z}\,H_{\alpha}\,\left(AA^{\dag}\right)_{\alpha\beta}\,H_{\beta} 
\,.
\end{equation}
Note that microscopic pairing theory implies $g_z > 0$~\cite{thu87}, in which case there is a cost in Zeeman energy for $S=1,M_s=0$ triplet pairs projected along $\vH$.

Spatial variations of the order parameter also incur a cost in kinetic and bending energies described by the gradient terms,
\begin{equation}\label{eq-gradient_energy}
f_\mathrm{grad}[A] 
\ns=\ns K_{1} A_{\alpha j,k}^{*} A_{\alpha j,k} 
\ns+\ns K_{2} A_{\alpha j,j}^{*} A_{\alpha k,k} 
\ns+\ns K_{3} A_{\alpha j,k}^{*} A_{\alpha k,j}
\,,
\end{equation}
where $A_{\alpha i,j}\equiv\grad_j A_{\alpha i}$. The gradient energies and related currents are discussed in more detail in Sec.~\ref{sec-Mass_Currents}.

The nuclear magnetic dipole-dipole interaction energy per atom is of order, $n\,(\gamma\hbar)^2\sim 10^{-4}\,\mbox{mK}$. This is a very weak perturbation compared to the binding pairing energy of Cooper pairs of order, $T_c\sim 1\,\mbox{mK}$. Nevertheless, the dipolar energy plays a central role in the NMR spectroscopy of the superfluid phases of \He, and specifically the spectroscopy of the vortex phases of rotating \Heb, because the dipole energy couples the spin and orbital degrees of freedom of the spin-triplet, p-wave condensate. 
Thus, in addition to the primary contributions to the GL functional (Eqs.~\ref{eq-condensation_energy}-\ref{eq-gradient_energy}), the mean nuclear dipole-dipole interaction energy, contributes to the GL functional a term second-order in the order parameter,
\be\label{eq-dipole_energy}
f_{\mathrm{dipole}} = g_D\,\left[|\Tr{A}|^2+\Tr{AA^*}-\nicefrac{2}{3}\Tr{AA^\dagger}\right]
\,,
\ee
where the material parameter, $g_D$, is determined by measurement of the slope of the square of the longitudinal resonance frequency, $\omega_{\text{B}}$, for bulk \Heb\, $g_D=\nicefrac{3}{5}\beta_{\text{B}}(1+F_0^a)^{-1}\,T_c(d(\hbar\omega_B)^2/dT\vert_{T_c})$~\cite{thu87}~\footnote{For a detailed analysis of the determination of $g_D$ see Ref.~\cite{thu01}.}, where $F_0^a$ is the exchange interaction for normal-state quasiparticles in units of the Fermi energy per atom of \He, and $\beta_{\text{B}}$ determines the bulk order parameter of \Heb\ (c.f. Sec.~\ref{sec-Material_Parameters}). The nuclear dipole energy is too weak to affect the relative stability of the vortex phases. But, when treated perturbatively, describes the dipolar energy of textures in rotating \Heb\ that are modified by the vortex currents and the intrinsic magnetization generated by rotation. These hydrodynamic and hydromagnetic effects are discussed in detail in Refs.~\cite{hak83,thu87}. Here we are interested in the internal structure and stability of the vortices in rotating \Heb, and thus we can neglect the nuclear dipole energy in our analysis of the energetics of the vortex phases.

\vspace*{-3mm}
\subsection{Material Parameters}\label{sec-Material_Parameters}
\vspace*{-3mm}

The material parameters, $\alpha$, $\{\beta_i|\,i=1\ldots 5\}$, $g_z$, and $\{K_a|\, a=1,2,3\}$ multiplying the invariants defining the GL functional, which in general are functions of temperature and pressure, are determined by the microscopic pairing theory for \He~\cite{ser83}. The coefficient of the second-order invariant determines the zero-field superfluid transition~\cite{thu87,wim15}$^{,}$\footnote{There is a very small correction to $\partial\alpha/\partial T|_{T_c}$ from the finite lifetime of quasiparticles which has no role in the relative stability of the vortex phases.},
\be
\alpha(T)=\nicefrac{1}{3}N_f\left(T/T_c-1\right)
\,,
\ee
where $N_f={m^* k_f}/{2\pi^2\hbar^2}$ is the single-spin normal-state density of states at the Fermi level expressed in terms of the quasiparticle effective mass, $m^*$, and Fermi wavenumber, $k_f$. The latter is determined by the particle density $n=k_f^3/3\pi^2$. In addition, the Fermi momentum, $p_f=\hbar k_f$, Fermi velocity, $v_f = p_f/m^*$ and Fermi energy, $E_f=\nicefrac{1}{2}v_f p_f$, determine the GL material parameters, all of which depend on pressure via the equilibrium particle density (see Table~\ref{table-material_parameters} of Appendix~\ref{app-material_parameters}).

For the homogeneous bulk phase it is convenient to represent the order paramter matrix in terms of an amplitude and normalized matrix, $A = \Delta\,a$ where $\Tr{aa^{\dag}} = 1$. Then for any stationary solution of the bulk free energy functional, Eq.~\ref{eq-GL_functional}, the pair density is $\Delta^2 = |\alpha(p,T)|/2\beta_a$, where $\beta_a$ is a local minimum of the functional $\beta[a]= \beta_{2}+\beta_{1}|\Tr{a a^{T}}|^2+\beta_{3}\Tr{aa^{T}(aa^{T})^*} +\beta_{4}\Tr{(a a^{\dagger})^2}+\beta_{5}\Tr{aa^{\dagger}(a a^{\dagger})^*}$. The corresponding bulk free energy density for the stationary solution is then, $f_a = \nicefrac{1}{2}\alpha\Delta_a^2=-\nicefrac{1}{4}\alpha^2/\beta_a$.

In weak-coupling BCS theory the \emph{relative} values of the five fourth-order materials parameters 
are uniquely determined,
\be\label{eq-beta_weak-coupling}
2\beta^{\text{wc}}_1=-\beta^{\text{wc}}_2=-\beta^{\text{wc}}_3=-\beta^{\text{wc}}_4=\beta^{\text{wc}}_5
\,,
\ee

\vspace*{-5mm}

\be
\mbox{where}\quad
\beta_1^{\text{wc}}=-\frac{7N_f\zeta(3)}{240 (\pi k_BT_c)^2}
\,.\hspace{30mm}
\ee

As a result the weak-coupling BCS formulation of GL theory predicts a unique bulk phase, the Balian-Werthamer (BW) state~\cite{bal63} defined by $A^{\text{B}}_{\alpha i}=\Delta_{\text{B}}\,\delta_{\alpha i}/\sqrt{3}$ where $\Delta_{\text{B}} = \sqrt{|\alpha(T)|/2\beta_{\text{B}}}$, which is the ground state at all pressures in zero magnetic field. The magnitude of the B-phase order parameter is defined by $\beta_{\text{B}}\equiv\beta_{12}+\onethird\beta_{345}$ where $\beta_{ijk\ldots}=\beta_{i}+\beta_{j}+\beta_{k}+\ldots$. In the weak-coupling theory $\beta_{\text{B}}^{\text{wc}}=\nicefrac{5}{3}|\beta^{\text{wc}}_1|$.
For comparison, the bulk A phase, first discussed as a possible ground state of \He\ by Anderson and Morel (AM)~\cite{and61}, is defined by $A^{\text{A}}_{\alpha i}=\Delta_{\text{A}}\,\hat\vz_{\alpha}\,\left(\hat\vx_{i} + i \hat\vy_{i}\right)/\sqrt{2}$, where $\Delta_{\text{A}}=\sqrt{|\alpha(T)|/2\beta_{\text{A}}}$ with $\beta_{\text{A}}\equiv\beta_{245}$, which in the weak-coupling limit becomes $\beta_{\text{A}}^{\text{wc}} = 2|\beta_{1}^{\text{wc}}|$. Thus, in weak-coupling theory the A phase is never stable relative to the B phase.

For inhomogeneous states the coefficients of the gradient energies determine the response of the order parameter to strong perturbations, e.g. the spatial variations, both suppression and growth, of order parameter components in the cores of vortices and topological defects. In the weak-coupling limit the stiffness coefficents are all given by
\ber
K^{\text{wc}}_1=K^{\text{wc}}_2=K^{\text{wc}}_3 = \frac{7\zeta(3)}{60}N_f\xi^2_0
\,,
\label{eq-GL_gradient}
\eer
where $\xi_0=\hbar v_f/2\pi k_BT_c$ is the Cooper pair correlation length in the $T=0$ limit. At temperatures close to $T_c$ the correlation length for spatial variations of the order parameter is given by the GL coherence length, 
\be\label{eq-GL_coherence_length}
\xi = \sqrt{\frac{K_1}{|\alpha(p,T)|}} 
    =\frac{\xi_{\text{GL}}}{\left(1-T/T_c\right)^{\nicefrac{1}{2}}}
\,,
\ee
where $\xi_{\text{GL}}=(7\zeta(3)/20)^{\nicefrac{1}{2}}\,\xi_0$ in the weak-coupling theory for the gradient energies.

The strength of the quadratic Zeeman energy for spin-triplet pairing is given by 
\ber
g^{\text{wc}}_z=\frac{7\zeta(3)}{48\pi^2}\frac{N_f\,(\gamma\hbar)^2}{\left[(1+F^a_0)\kb T_c\right]^2}
\,,
\eer
where $\gamma$ is the nuclear gyromagnetic ratio for the \He\ nucleus and $F_0^a$ is the exchange interaction. The latter is ferromagnetic, varying from $F_0^a=-0.723$ at $p=0$ bar to $F_0^a=-0.778$ at melting pressure, $p=34$ bar. Thus, combined with the large effective mass at high pressures the nuclear magnetic susceptibility is enhanced by an order of magnitude relative to the Pauli susceptibility at the same density. This enhancement was the basis for ferromagnetic spin-fluctuation exchange models for the superfluid transition to spin-triplet pairing~\cite{lay71}. For convenience we include all relevant material parameters as a function of pressure, with references to measured values, in Appendix~\ref{app-material_parameters}.

\subsection{Strong-Coupling Theory}\label{sec-Strong_Coupling_Theory}

A strong-coupling formulation of GL theory that accounts for the relative stability of the bulk A- and B phases, and specifically the bulk A-B transition line, $T_{\text{AB}}(p)$ for pressures above the polycritical point, $p\gtrsim \pPCP$ was introduced in Ref.~\cite{wim15}. This strong-coupling GL functional is defined by the fourth-order GL material parameters,

\vspace*{-3mm}

\be\label{eq-strong-coupling-betas}
\beta_i(p,T)=\betawc{i}(p)+\frac{T}{T_c}\betasc{i}(p)
\,.
\ee
The weak-coupling parameters, $\beta_i^{\text{wc}}(p)$, are obtained from the leading order contribution to the Luttinger-Ward free-energy functional as an expansion in the small parameter $T_c/T_f$, where $T_f\approx 1\,\mbox{K}$ is the Fermi temperature. The $\beta_i^{\text{wc}}(p)$ are expressed in terms of pressure-dependent material parameters as shown in Eq.~\ref{eq-beta_weak-coupling}, and can be calculated from the material parameters provided in Table \ref{table-material_parameters} of Appendix~\ref{app-material_parameters}.

The next-to-leading order corrections to the weak-coupling GL functional enter as corrections to the fourth-order weak-coupling material coefficients. These terms are of order $\Delta\betasc{i}\approx\betawc{i}(T/T_f)\langle w_{i}|\mathsf T|^2\rangle$, where $\langle w_{i}|\mathsf T|^2\rangle$ is a weighted average of the square of the scattering amplitude for binary collisions between quasiparticles on the Fermi surface~\cite{rai76}.
At high pressures, strong scattering of quasiparticles by long-lived spin fluctuations largely compensates the small parameter $T/T_f$, resulting in substantial strong-coupling corrections to the weak-coupling theory, and the stabilization of the AM state as the A phase~\cite{sau81b}.

In the analysis of the stability of the vortex phases of \Heb\ we use improved results for the strong-coupling parameters based on a recent determination of the effective interactions and scattering amplitudes that account for the body of normal-state thermodynamic and transport data on 
liquid \He\ over the full pressure range below the melting pressure, as well as the heat capacity jumps for the bulk A and B phases at $T_c(p)$ in zero field~\cite{wim19}. The results of this analysis provide a quantitative theory for the thermodynamic properties of the bulk A and B phases of superfluid \He\ at all pressures, including a quantitative determination of the bulk A-B transition line, $\TAB(p)$, for pressures above the polycritial point, $\pPCP{}$, as well as the temperature dependence of the free energy, entropy and heat capacity at all temperatures below $T_c$.
The strong-coupling corrections to the $\beta$-parameters obtained from microscopic theory~\cite{wim19}, listed in Table~\ref{table-material_parameters} of Appendix~\ref{app-material_parameters}, reproduce the heat capacity jumps for the A and B transitions over the full pressure range. In particular, the A phase correctly appears as a stable phase above the polycritical point $\pPCP{}=21.22\mbar{}$.
However, in the standard formulation of the GL theory in which the $\beta_i$ parameters are evaluated at $T_c$, and thus treated as functions only of pressure, the A phase is the only stable phase for all temperatures and pressures above $\pPCP{}$, i.e. the standard fourth-order GL theory fails to account 
for the bulk A-B transition at $\TAB{}(p)$.

In Ref.~\cite{wim15} the missing A-B transition line was traced to the omission of the temperature dependence of the fourth-order $\beta$ parameters in the neighborhood of a triple point.
The latter is defined by the intersection of the second-order transition line given by $\alpha(T_c,p) = 0$, and the first-order boundary line separating the A- and B-phases given by $\Delta\beta_{AB}(\TAB{},p)\equiv\beta_{A}-\beta_{B}=0$, where $\beta_{A}\equiv\beta_{245}$ and $\beta_{B}\equiv\beta_{12}+\nicefrac{1}{3}\beta_{345}$. At the PCP we have $\TAB{}(\pPCP{})=T_c(\pPCP{})$. But, for $p>\pPCP{}$ the lines separate and we must retain both the temperature and pressure dependences of $\Delta\beta_{AB}(T,p)$ to account for $\TAB{}(p)$ in the vicinity of $\pPCP{}$. 
The degeneracy between the A- and B-phases near $\pPCP{}$ is resolved by retaining the linear $T$ dependence of the strong-coupling corrections to the $\beta$ parameters.
The suppression of the strong-coupling terms originates from the reduction in phase space for quasiparticle scattering with decreasing temperatures, and is the basis for the temperature scaling of the strong-coupling corrections in Eq.~\ref{eq-strong-coupling-betas}.
The analysis and predictions for the vortex phases of superfluid \He{} reported here are based on the strong-coupling material parameters calculated and reported in Ref.~\cite{wim19}, combined with the known pressure-dependent material parameters, $m^*$, $v_f$, $T_c$, and $\xi_0$ as listed in Table \ref{table-material_parameters} in Appendix~\ref{app-material_parameters}, and the temperature scaling in Eq.~\ref{eq-strong-coupling-betas} that accounts for the reduction in strong-coupling effects below $T_c$. 
The resulting bulk phase diagram predicted by strong-coupling GL theory accounts remarkably well for the experimental A-B transition line, $\TAB{}(p)$, as shown in Fig. \ref{fig-Vortex_Phase_Diagram}, as well as the heat capacity jumps of the bulk A and B phases.
We emphasize that the predictions of the relative stability of the A and B phases by the strong-coupling GL functional is validated by microscopic calculations of $T_{\text{AB}}(p)$~\cite{wim19} based on the formulation of the strong-coupling theory developed in Refs.~\cite{rai76,sau81b,sau81c,ser83}.

%--------------------------------------------------------------------------------------
\vspace*{-3mm}
\section{Vortex States in Superfluid \Heb}\label{sec-Vortex_States}
\vspace*{-3mm}

For rotating equilibrium of superfluid \Heb\ the inter-vortex spacing for singly-quantized, axially symmetric vortices organized on a hexagonal lattice is $d=\left(\kappa/\sqrt{3}\Omega\right)^{\rfrac{1}{2}}$. For an angular velocity of $\Omega=1.7\,\mbox{rad/s}$ the vortex unit cell dimension is $d=0.150\,\mbox{mm}\approx 6.7\times 10^{3}\xi_0$ at $p=18\mbox{bar}$. Thus, most of the vortex unit cell is occupied by a texture of the bulk B phase,
\be\label{eq-OP_B-phase_texture}
A_{\alpha i}(\vr)=\Delta_{\text{B}}\,R_{\alpha i}[\hat\vn,\vartheta]\,e^{i\Phi}
\,,
\ee
where $R_{\alpha i}[\hat\vn,\vartheta]$ is an orthogonal matrix that defines the relative angle of rotation, $\vartheta$, about the local axis $\hat\vn$, between the spin- and orbital coordinates of the Cooper pairs. 

The texture, $\hat\vn(\vr)$, is determined by a competition of surface and bulk nuclear dipolar enegies, modified by the pair-breaking and orienting effects of the vortex flow and the intrinsic vortex magnetization. These textural energies are treated perturbatively after the vortex structure is calculated for a fixed choice of the relative orientation of the spin and orbital coordinates of the Cooper pairs~\cite{thu86,thu87}.
In particular, we can neglect the nuclear dipole energy for distances $r < \xi_{\text{D}}=\sqrt{K_1/g_{D}} \simeq 15\,\mu\mbox{m}\approx 6.7\times 10^2\,\xi_0$ at $p=18\,\mbox{bar}$. Thus, we can choose a convenient computational cell dimension $\xi_0\ll d_c\ll \xi_{\text{D}}$ which allows a converged solution at distances well beyond the vortex core, but still at distances well within the dipole coherence length. Thus we can omit the dipole energy and work in a convenient spin- and orbital coordinate system. 
We use the basis of aligned spin and orbital coordinates to determine the vortex structures and free energy of the vortex states, and in the calculations reported here the computational cell dimension is $d_c=60\xi$, where $\xi$ is the temperature-dependent coherence length defined in Eq.~\ref{eq-GL_coherence_length}.

%\eject

%----------------------------------------------
\onecolumngrid
\begin{figure}[t]
\begin{minipage}{\textwidth}
\hspace*{-12mm}
\begin{tabular}{ccc}
\raisebox{-0.5\height}{\includegraphics[height=7.175cm]{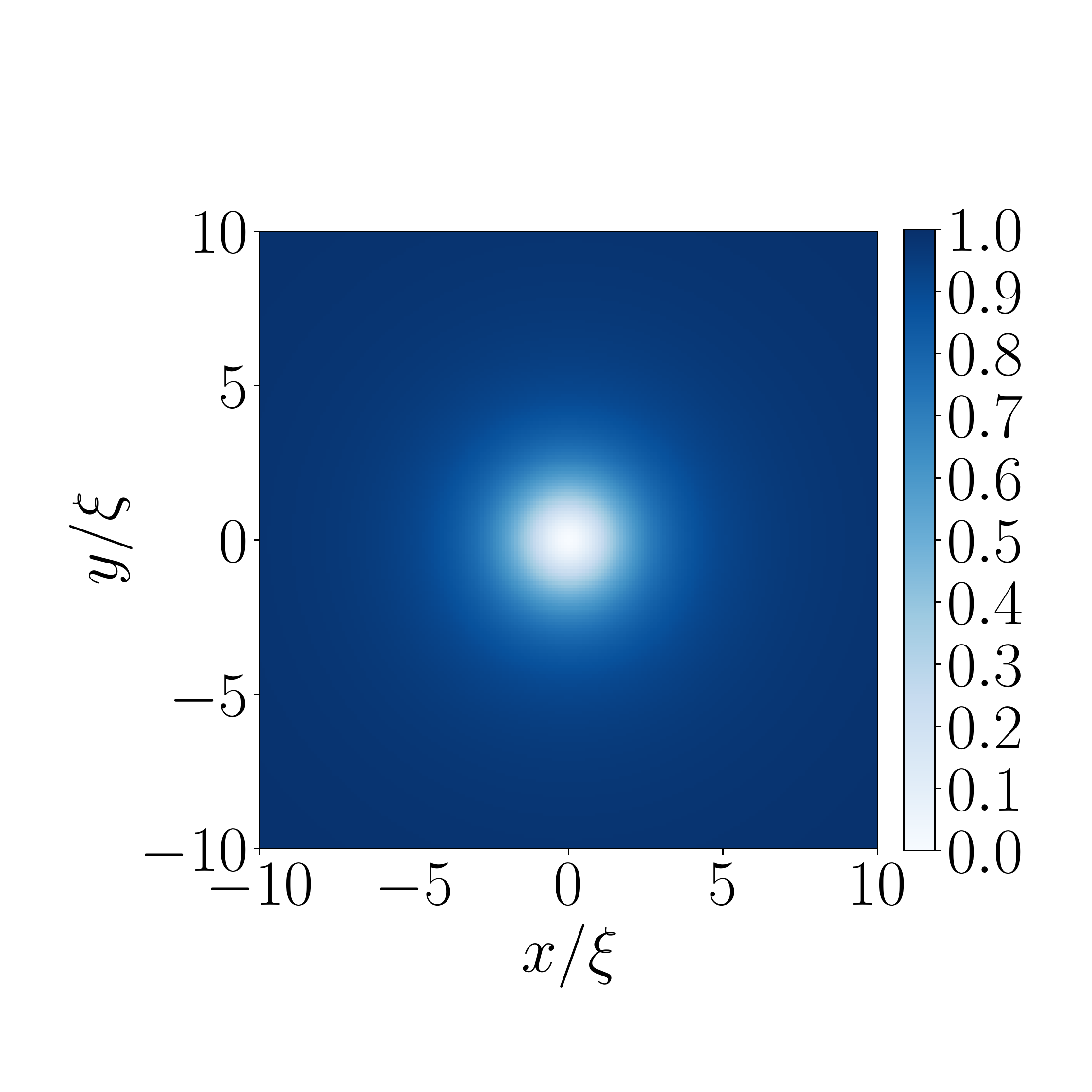}}
&
\hspace*{-5mm}
\raisebox{-0.5\height}{\includegraphics[height=6.250cm]{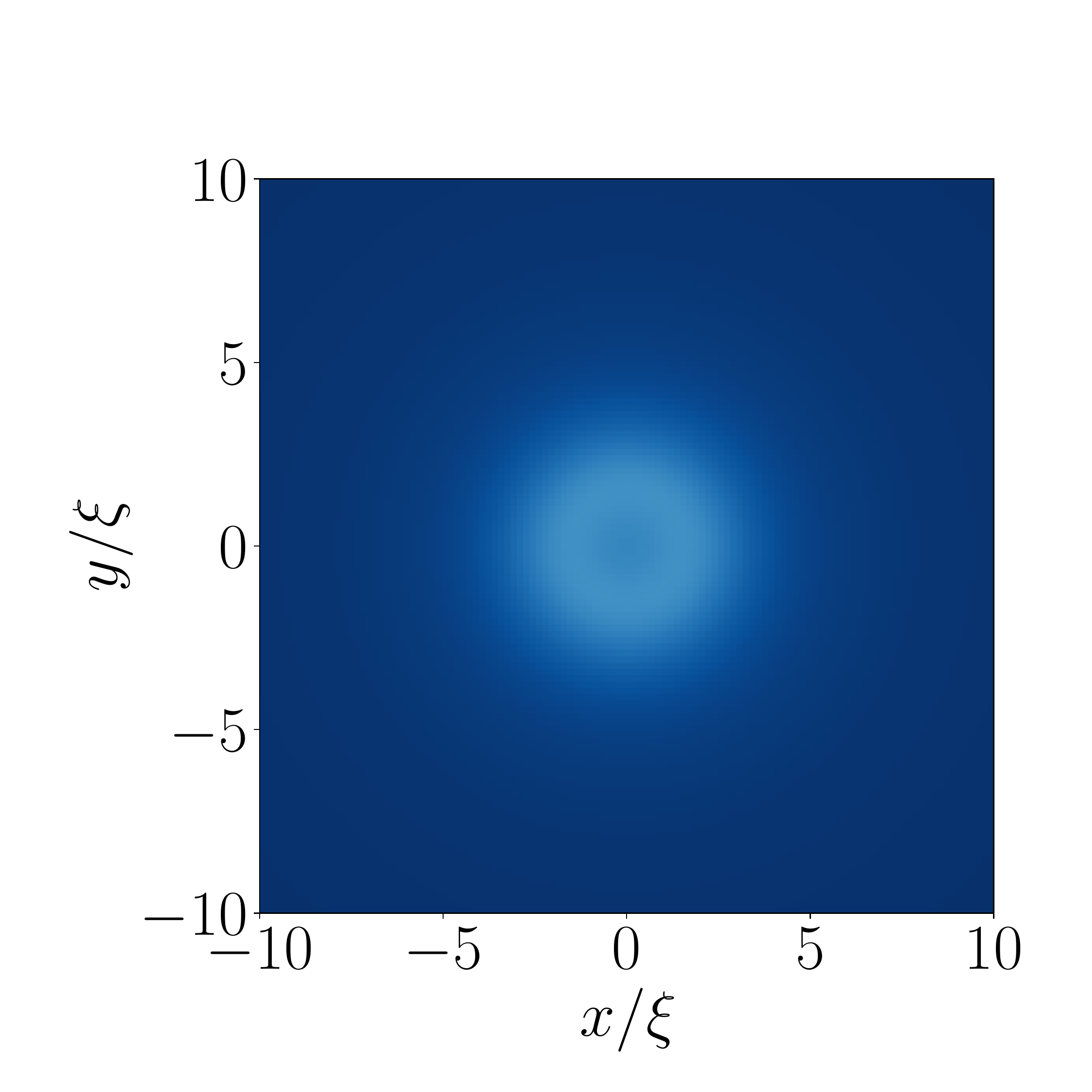}}
&
\hspace*{-5mm}
\raisebox{-0.5\height}{\includegraphics[height=6.250cm]{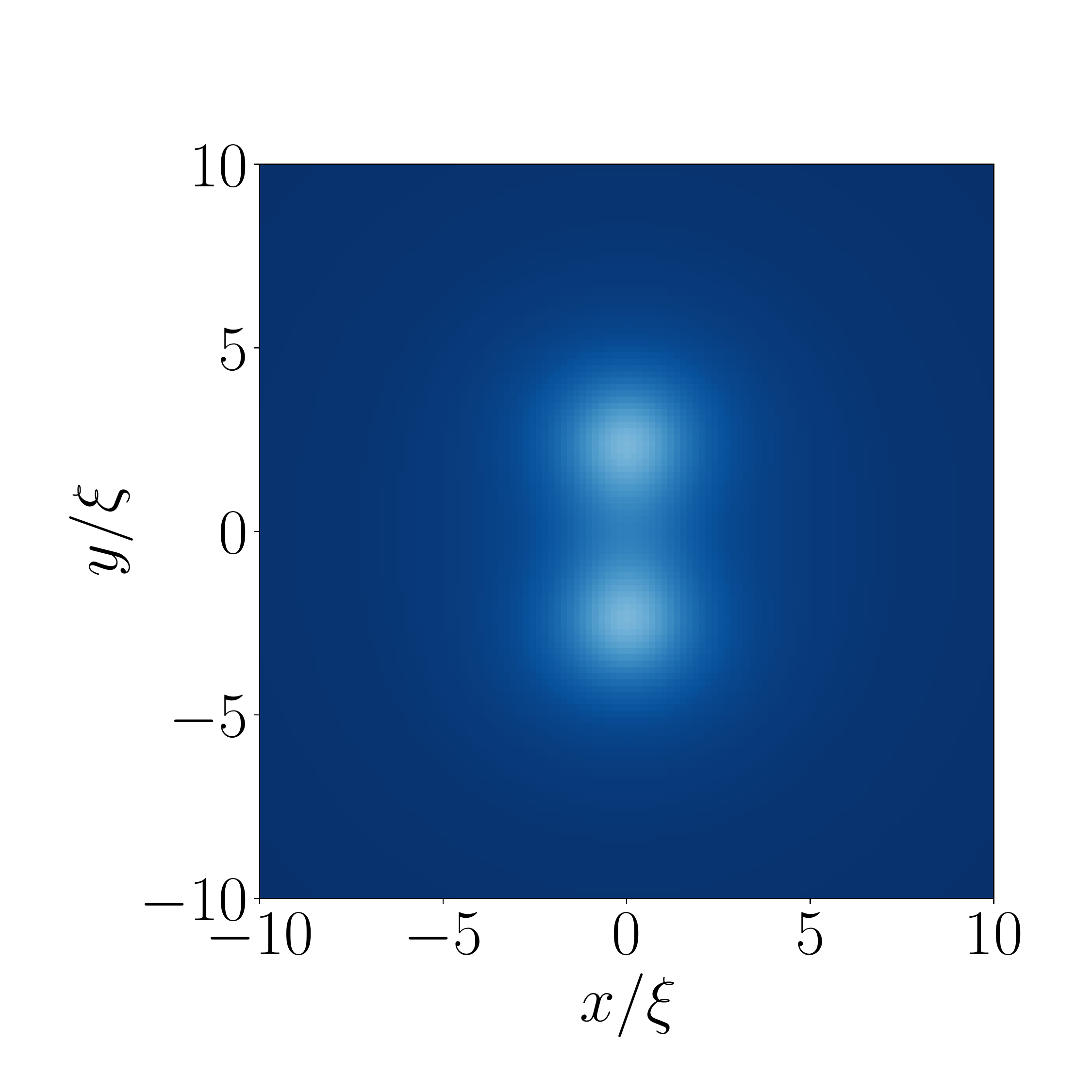}}
\end{tabular}
\caption{
Left panel: The axially symmetric B-phase vortex (``o-vortex'') has a hard core with a node 
in $\Delta(\vr)$.
Center panel: The axially symmetric A-core vortex (``v-vortex'') has a suppressed, but non-vanishing, condensate density in the core which is predominantly the order parameter for the bulk A phase. 
Right panel: The D-core vortex has a ``double core'' structure that spontaneously breaks axial rotation symmetry.
All plots are of the condensate density, $|\Delta(\vr)|^2\equiv\Tr{AA^{\dag}}$, in units of that for the bulk B-phase, $\Delta_{\text{B}}^2\equiv\Tr{A_{\text{B}}A_{\text{B}}^{\dag}}$. 
The solutions of the GL equations for the o-vortex, A-core and D-core vortices correspond to 
$p=10$ bar and $T=0.25T_c$, 
$p=34$ bar and $T=0.75T_c$, 
$p=20$ bar and $T=0.55T_c$, respectively. 
}
\label{fig-OP_Density}
\end{minipage}
\end{figure}
\twocolumngrid
%----------------------------------------------

\vspace*{-7mm}
\subsection{Euler-Lagrange Equations}\label{sec_Euler_Lagrange_Equations}
\vspace*{-3mm}

To determine equilibrium and metastable vortex phases we obtain stationary solutions of the strong-coupling GL functional, $\cF[A]$, defined by Eqs.~\ref{eq-GL_functional}-\ref{eq-gradient_energy}. The equilibrium and metastable states in zero field are solutions of the Euler-Lagrange equations of $\cF[A]$ defined by the functional gradient, $G[A]\equiv\delta\cF/\delta A^\dagger=0$, 
\ber\label{eq-GL_equations}
&&
-
\alpha(T) A_{\alpha i}
+
K_1\grad^2 A_{\alpha i}
+
(K_{2}+K_{3})\grad_i\grad_j A_{\alpha j}
\nonumber\\
&&
-2\left[
\beta_1 A^*_{\alpha i}\Tr{AA^{T}}
+
\beta_2 A_{\alpha i} \Tr{AA^\dagger}
+
\beta_3 (A A^{T}A^*)_{\alpha i}
\right.
\nonumber\\
&&
\left.
+
\quad
\beta_4 (A A^\dagger A)_{\alpha i}
+
\beta_5 (A^*A^{T}A)_{\alpha i}
\right]
=0 
\,.
\eer
In zero magnetic field, at distances far from the core of a quantized vortex, $|\vr|\gg\xi$, the order parameter approaches the bulk B phase order parameter with a global phase that reflects the topological winding number of the vortex,
\be\label{eq-OP_boundary_condition}
A_{\alpha i}(\vr)\xrightarrow[\vr\rightarrow\vr_c]{}\frac{\Delta_{\text{B}}}{\sqrt{3}}\,\delta_{\alpha i}\,e^{i\Phi(\vr)}
\,,
\ee
where $\Phi(\vr)$ is constrained by phase quantization, $\oint \grad\Phi\cdot d\vell=p\,2\pi$ with $p\in\{0,\pm 1,\pm 2,\ldots\}$, which we enforce with $\Phi(\vr)=p\,\phi$ imposed on the computational boundary, where $\phi$ is the azimuthal angle in cylindrical coordinates defined with respect to the phase singularity.
 
For external fields parallel to the axis of rotation, $\vH=H\hat\vz$, we must add a term representing the Zeeman energy, $-g_z H^2\,\delta_{\alpha z}\,A_{zi}$, to the left side of Eqs.\ref{eq-GL_equations}.
In an external magnetic field we must also modify the boundary condition to incorporate gap distortion by the Zeeman energy on the bulk B phase order parameter. The boundary condition in Eq.~\ref{eq-OP_boundary_condition} is replaced by
\be\label{eq-boundary_condition_field}
A_{\alpha i}(\vr)\xrightarrow[\vr\rightarrow\vr_c]{}
\nicefrac{1}{\sqrt{3}}
\left[
\Delta_{\perp}(\delta_{\alpha i}-\hat\vz_{\alpha}\hat\vz_{i})
+
\Delta_{\parallel}\hat\vz_{\alpha}\hat\vz_{i}
\right]\,e^{i\Phi(\vr)}
\,,
\ee
where the field-induced gap distortion of the order parameter is given by
\ber
\Delta_{\perp}&=&\Delta_{\text{B}}
\sqrt{1+\frac{\beta_{12}}{\beta_{345}} \frac{H^2}{H_0^2} }
\,,
\\
\Delta_{\parallel}&=&\Delta_{\text{B}}
\sqrt{1-\frac{2\beta_{12}+\beta_{345}}{\beta_{345}}\frac{H^2}{H_0^2} }
\,,
\eer
where $H_0\equiv \sqrt{|\alpha(p,T)|/g_z}$ is the field scale at which the bulk B phase is strongly deformed or destroyed.

In order to obtain stationary state solutions to the GL equations a simple method is to find a solution of the discretized time-dependent GL equation~\cite{thu87},
\be\label{eq-TDGL_equation}
\pder{A_{\alpha i}}{t} 
=-\Gamma\,\frac{\delta\cF}{\delta A_{\alpha i}^{*}} \equiv -\Gamma\,G[A]_{\alpha i}
\,,
\ee
which relaxes to a stationary state satisfying Eq.~\ref{eq-GL_equations}, i.e. $G[A]_{\alpha i}=0$. Here we use the quasi-Newton, Limited-memory Broyden–Fletcher–Goldfarb–Shanno algorithm (L-BFGS)~\cite{nocedal1980,noc18} to obtain stationary state solutions of $G[A]=0$ that is far more efficient than relaxation based on Eq.~\ref{eq-TDGL_equation}. Our implementation of the L-BFGS algorithm is outlined in App.~\ref{app-l-bfgs} where we also provide a benchmark comparison of the improvement in rate of convergence to a solution of the GL equations using the L-BFGS algorithm compared to relaxation~\footnote{See also Ref.~\cite{vil02} for implementations of gradient decent algorithms for solving the GL equations.}.
In App.~\ref{app-benchmark-GL} we provide a benchmark of our GL equation solver.

In general there are multiple stationary-state solutions to Eq.~\ref{eq-GL_equations}. As a result convergence to a steady-state solution can also be influenced by the initialization of the order parameter.
Thus, in addition to the boundary condition at the edge of the computational cell, we use targeted initialization of the order parameter to find stationary states with different symmetries. The free energy of the converged stationary solutions determines the equilibrium phase.   
For example, to obtain a stationary solution for the D-core vortex, either equilibrium or metastable, a non-axi-symmetric initialization of the order parameter is used which converges to the targeted vortex efficiently. If the targeted vortex state is not a local minimum then symmetry breaking at the initialization stage will not yield a vortex with that broken symmetry. 

Our analysis based on the strong-coupling free energy functional identifies the three stationary state vortex solutions for \Heb\ in zero magnetic field, originally discussed by Ohmi et al.~\cite{ohm83} (o-vortex), Salomaa and Volovik~\cite{sal83} (A-core vortex) and Thuneberg~\cite{thu86} (D-core vortex). Figure \ref{fig-OP_Density} illustrates the basic structure of these three vortices in terms of their condensate densities. The o-vortex is ``singular'' with condensate density vanishing at the vortex core center. The A-core vortex has a ``superfluid core'' with finite condensate density in the core. The D-core vortex breaks rotational symmetry exhibiting a double core structure, also with a finite condensate density.

%-------------------------------------------------
\begin{figure}[t] 
\centering
\includegraphics[width=\columnwidth]{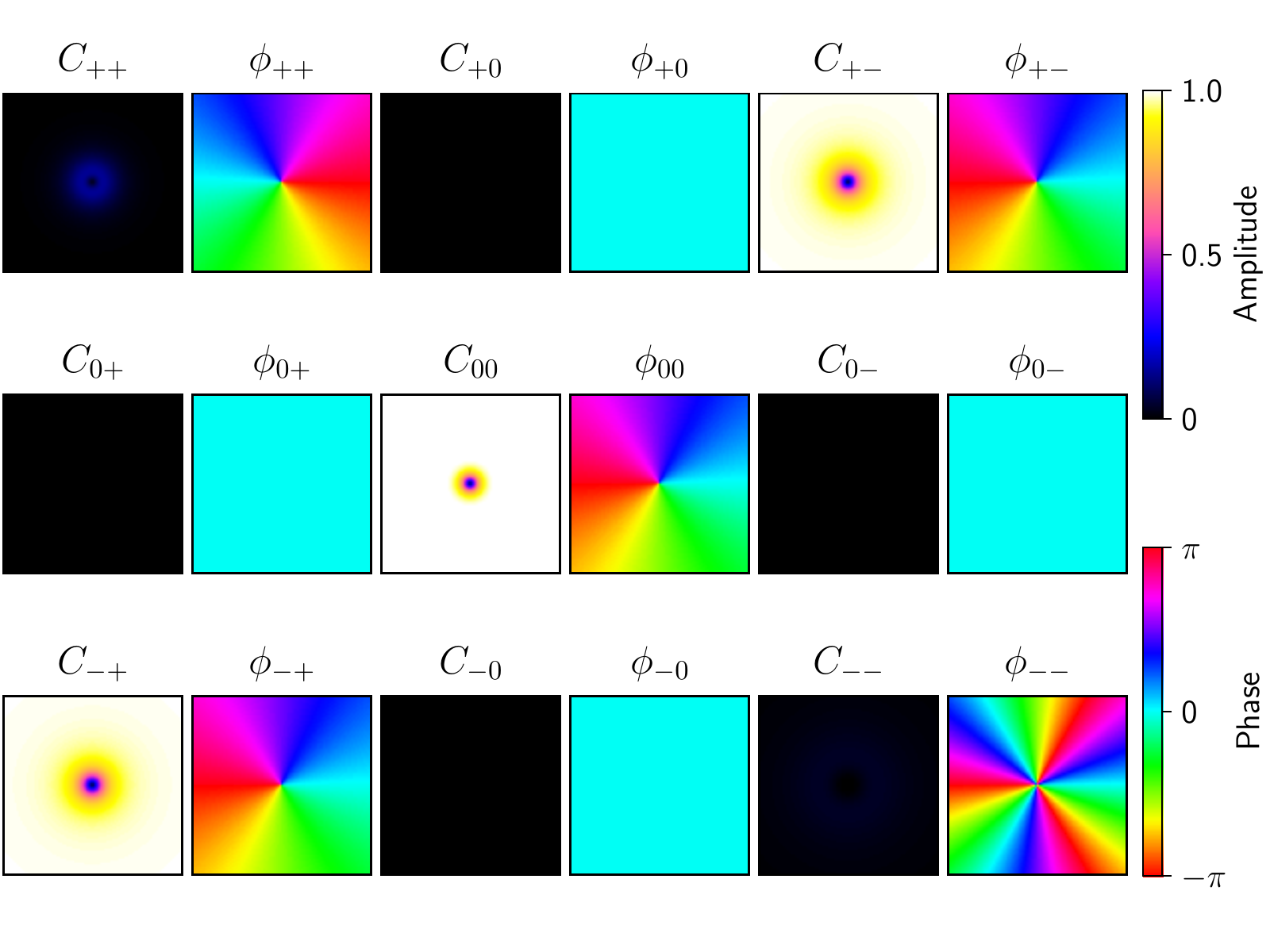}
\caption{
Amplitudes and phases of the components of the o-vortex at $p=10$ bar and $T=0.25\,T_c$, shown on a square grid with $x$ and $y$ ranging from $[-5\,\xi,+5\,\xi]$. The computational grid was $60\,\xi\times 60\,\xi$ with grid spacing $h=0.1\xi$.
The o-vortex retains the maximal symmetry of the stationary vortex solutions for the B-phase; axial rotation symmetry and time-reversal symmetry are preserved. Amplitudes with $N=0$ and $N=2$ vanish by symmetry. Thus, the o-vortex has vanishing condensate density in the core. 
}
\label{fig-NCore_AmpPhase}
\end{figure}
%-------------------------------------------------

\vspace*{-5mm}
\subsubsection*{Initialization \& Soft Modes of the Order Parameter}
\vspace*{-3mm}

The stationary ``o-vortex'' is obtained by initialization of the order parameter as a singly-quantized local B-phase vortex of the form $A_{\alpha i}(\vr)=\nicefrac{1}{\sqrt{3}}\Delta_{\text{B}}\,\tanh(|\vr|/\sqrt{2}\xi)\,\delta_{\alpha i}\,\exp{(i\phi)}$.
However, to target vortices with lower symmetry we need to break additional symmetries, and it is useful to identify the soft modes of the order parameter associated with relative spin-orbit rotation symmetry of bulk \Heb.

In the absence of boundaries, magnetic fields, rotation and neglecting the nuclear dipole energy the  
bulk B phase order parameter has a large degeneracy space associated with relative spin-orbit rotations described by the rotation matrix $R_{\alpha i}[\hat\vn,\vartheta]$, which defines the orientation of the spin coordinates of the Cooper pairs relative to the orbital coordinate axes. 

For the analysis of the internal structure of vortex states in rotating \Heb, and their relative energies, the spin-orbit rotational degeneracy is partially resolved by the vortex flow. At distance scales $\xi \ll r \ll \xi_D$ the spin-orbit rotational degeneracy is a soft mode leading to some amplitudes of the order parameter developing long-range, power-law tails $\propto 1/r, 1/r^2$. The slow spatial variations of these modes for vortices in \Heb\ is discussed in detail in Refs.~\cite{ohm83,has85,thu87,sil15}.

In our analysis we use the asymptotic behavior of the soft-modes to target specific stationary vortex solutions. For the A-core vortex, we initialize the components, $A_{xz}$ and $A_{zx}$, to vary as $1/r$ and components, $A_{xy}$ and $A_{yx}$, to vary as $1/r^2$ for $|\vr| > 5\xi$.
To target the D-core vortex, we initialize by breaking axial symmetry by introducing a change in sign between the $A_{xz}$ and $A_{yz}$ amplitudes, \emph{and} seed the cores of the amplitudes with $4\pi$ phase winding (c.f. Eq.~\ref{eq-OP_Lz+Sz_basis} in Sec.~\ref{sec-axial_symmetric_vortices}) by initializing $A_{xz}$, $A_{zx}$, $A_{yz}$, and $A_{zy}$ with non-zero values in a small region of the core near $\vr=(0,0)$.

%----------------------------------------------------------------------------------------
\vspace*{-3mm}
\subsection{Axially Symmetric Vortex States in \Heb}\label{sec-axial_symmetric_vortices}
\vspace*{-3mm}

For axially symmetric, singly-quantized vortices, the circulation of each vortex in the asymptotic limit, $|\vr|\gg\xi$, $\oint\grad\Phi\cdot d\vell=2\pi$, is satisfied by $\Phi(\vr)=\phi$. The resulting mass current and moduli of the amplitudes for all components of the order parameter are axially symmetric. The simplest axially symmetric vortex is the local B-phase vortex first discussed by Ohmi et al.~\cite{ohm83}. 
The B phase of \He\ is invariant under joint spin and orbital rotations. Thus, for axially symmetric vortices, or vortices with weakly broken axial symmetry, it is instructive to represent the order parameter in the basis of angular momentum eigenvectors, $\{\pmb\lambda^{\mu}|\mu=-1,0,+1\}$, where $\pmb\lambda^0=\hat\vz$ and $\pmb\lambda^{\pm}=(\hat\vx\pm i\hat\vy)/\sqrt{2}$~\cite{ohm83}. These basis vectors satisfy the orthogonality relations, $\pmb\lambda^{\mu}\cdot\pmb\lambda^{\nu *} = \delta^{\mu\nu}$. We can transform a p-quantized vortex from the spin-orbit aligned Cartesian basis to the angular momentum basis by writing,
\be 
A_{\alpha i}(\vr)=\frac{\Delta_{\text{B}}}{\sqrt{3}}
\sum_{\mu,\nu}\,
\lambda_\alpha^\mu\,
\left[
A^{\mu \nu}(\vr)
\right]
\lambda_i^\nu
\,.
\ee
where $A^{\mu\nu}(\vr)\equiv C_{\mu \nu}(\vr)\,e^{iN{\mu\nu}\phi}$ are the complex order parameter amplitudes in the angular momentum basis, expressed in terms of amplitudes, $C_{\mu\nu}(\vr)$, and phases, $\phi_{\mu\nu}=N_{\mu\nu}\phi$. The $N_{\mu\nu}$ are integer winding numbers for the phase of the $\mu,\nu$ component. Asymptotically, for a p-quantized vortex of the B-phase
\be\label{eq-p_quantized_B_vortex}
A_{\alpha i}(\vr)\xrightarrow[|\vr|\rightarrow\infty]{}\frac{\Delta_{\text{B}}}{\sqrt{3}}
\delta_{\alpha i}\,
e^{ip\phi}
\,.
\ee

A p-quantized B phase vortex that is also axially symmetric is an eigenstate of the generator for
axial rotations, i.e.
\be\label{eq-axial_symmetric_vortex} 
J_z\,A_{\alpha i}(\vr) = j\hbar\,\,A_{\alpha i}(\vr)
\,,
\ee
with $j\in\{0,\pm 1,\pm 2,\ldots\}$. The total angular momentum projected along the axis of symmetry, $J_z= L_z^{\text{cm}} + L_z^{\text{int}} + S^{\text{int}}_z$, is the sum of the operator for the center-of-mass orbital angular momentum of the Cooper pairs, $L_z^{\text{cm}} = -i\hbar\partial_{\phi}$, and the internal orbital and spin angular momentum operators, $L_z^{\text{int}}$ and $S_z^{\text{int}}$. The latter yield,
\be
L_z^{\text{int}}\pmb\lambda^{\nu} = \nu\hbar\pmb\lambda^{\nu}
\,,\quad
S_z^{\text{int}}\pmb\lambda^{\mu} = \mu\hbar\pmb\lambda^{\mu}
\,.
\ee
The condition in Eq.~\ref{eq-axial_symmetric_vortex} must also apply to the asymptotic limit in Eq.~\ref{eq-p_quantized_B_vortex}, which requires $j=p$.
Imposing the axial symmetry condition, $J_z A_{\alpha i}(\vr) = p\,A_{\alpha i}(\vr)$, for any $|\vr|$ then fixes the phase of each $(\mu,\nu)$ component, $N_{\mu\nu}=p-\mu-\nu$. Thus, the form of the order parameter for a p-quantized, axially symmetric vortex becomes~\cite{ohm83},
\be 
A_{\alpha i}(\vr)=
\frac{\Delta_{\text{B}}}{\sqrt{3}}
\sum_{\mu,\nu}
\lambda_\alpha^\mu\,
\left[
C_{\mu \nu}(r)\,e^{i(p-\mu-\nu)\phi}
\right]\,
\lambda_i^\nu 
\,.
\ee
For a singly-quantized ($p=1$) B phase vortex we can organize the components into a matrix labeled by the orbital and spin angular momentum indices,
\be\label{eq-OP_Lz+Sz_basis}
\left[A^{\mu\nu}\right] = 
\left(
\begin{array}{lll}
C_{++}\,e^{-i\phi}
&
C_{+0}
&
C_{+-}\,e^{+i\phi}
\\
C_{0+}
&
C_{00}\,e^{+i\phi}
&
C_{0-}\,e^{+2i\phi}
\\
C_{-+}\,e^{+i\phi}
&
C_{-0}\,e^{+2i\phi}
&
C_{--}\,e^{+3i\phi}
\end{array}
\right)
\,.
\ee  

In Fig.~\ref{fig-NCore_AmpPhase} we show the amplitude and phase structure of a stationary solution of Eq.~\ref{eq-GL_equations} for the most symmetric singly quantized vortex state in \Heb. This is the ``o-vortex'', or ``normal-core vortex'',  which is  ``singular'' in the sense that all non-vanishing components incur a phase winding, and therefore force these amplitudes to vanish as $|\vr|\rightarrow 0$. This is clear from the results shown in Fig.~\ref{fig-NCore_AmpPhase} where the dominant components are $C_{+-}$, $C_{00}$ and $C_{-+}$, all of which vanish as $|\vr|\rightarrow 0$. Each of these dominant amplitudes have the same phase winding, $\phi_{+-}=\phi_{00}=\phi_{-+}=\phi$, as shown in the corresponding phase plots of Fig.~\ref{fig-NCore_AmpPhase}. 
In addition, the o-vortex develops very small sub-dominant amplitudes, $C_{++}$ and $C_{--}$, with phase windings of $N_{++}=-1$ and $N_{--}=+3$, respectively, also shown in Fig.~\ref{fig-NCore_AmpPhase}.

%-------------------------------------------------
\begin{figure}[t] 
\centering
\includegraphics[width=\columnwidth]{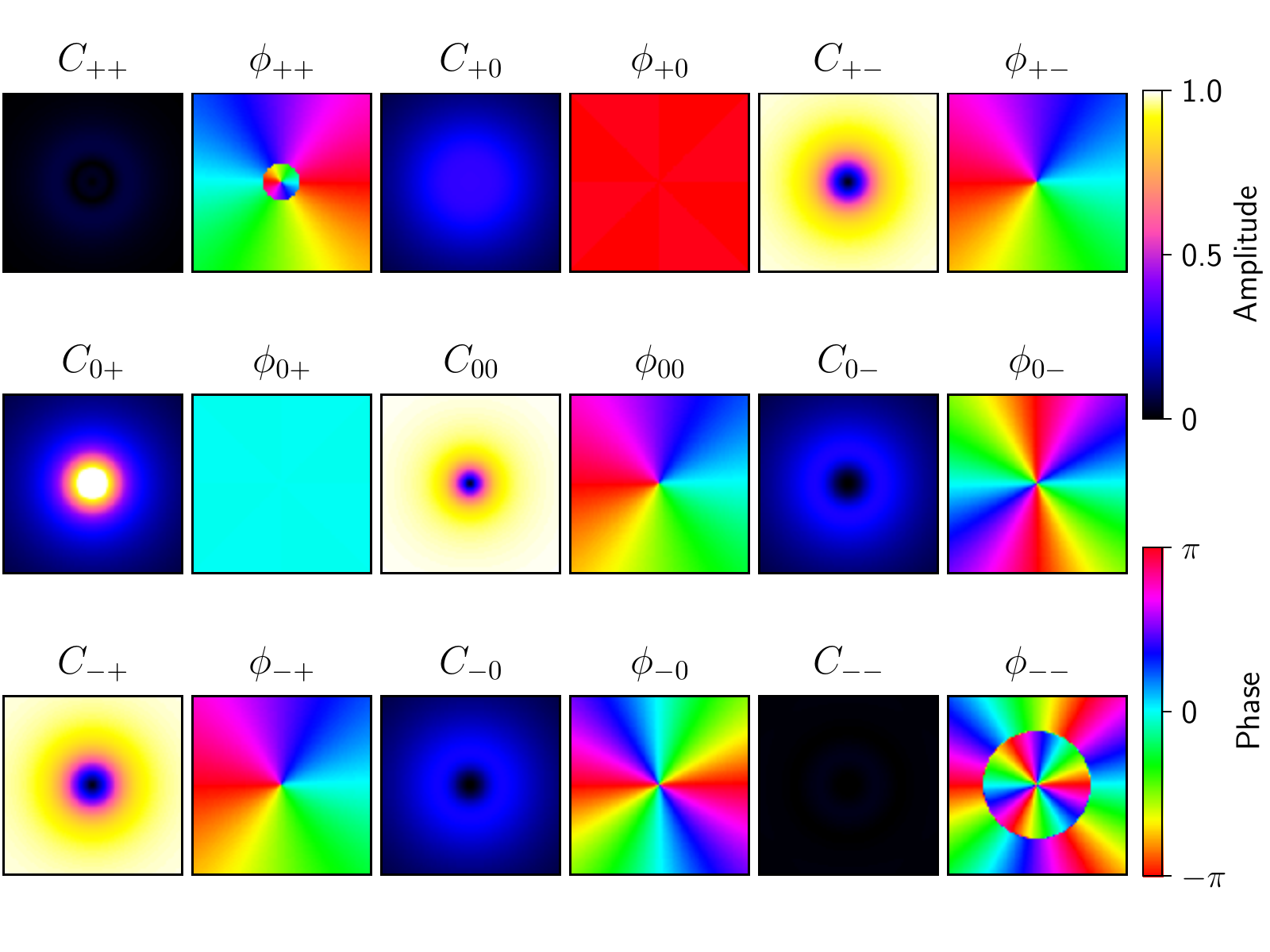}
\caption{
Amplitudes and phases of the components of the A-core vortex at $p=34$ bar and $T=0.75\,T_c$, shown and computed on the same grid as that in Fig.~\ref{fig-NCore_AmpPhase}. 
The key amplitudes defining the A-core vortex are the amplitudes with zero phase winding - the A phase $C_{0+}$ and the spin-polarized $\beta$ phase, $C_{+0}$. The amplitudes with winding number $N=2$, $C_{-0}$ and $C_{0-}$ are also important in terms of the relative stability between the A-core and D-core vortex states, as discussed in Sec.~\ref{sec-non-axial_symmetric_vortices}.
}
\label{fig-ACore_AmpPhase}
\end{figure}
%-------------------------------------------------

A key observation regarding the o-vortex is that the two amplitudes with zero phase winding, $C_{0+}$ and $C_{+0}$, are \emph{identically zero}. The amplitude $C_{0+}$ represents the equal-spin, chiral A phase with intrinsic angular momentum $J_z^{\text{int}}=+\hbar$ from the orbital state of the Cooper pairs, while $C_{+0}$ is the $\beta$-phase, also with $J_z^{\text{int}}=+\hbar$ from the spin state of the Cooper pairs. 
Components with zero phase winding can support finite amplitudes in the vortex core. This was the observation of Ref.~\cite{sal83}, and the basis for the prediction of a ferromagnetic vortex in which both amplitudes, $C_{0+}$ and $C_{+0}$, are finite in the core. This is the ``A-core'' vortex, which is a stationary solution of the GL equations (Eqs.~\ref{eq-GL_equations}).

%-------------------------------------------------
\begin{figure}[t]
\centering
\includegraphics[width=\columnwidth]{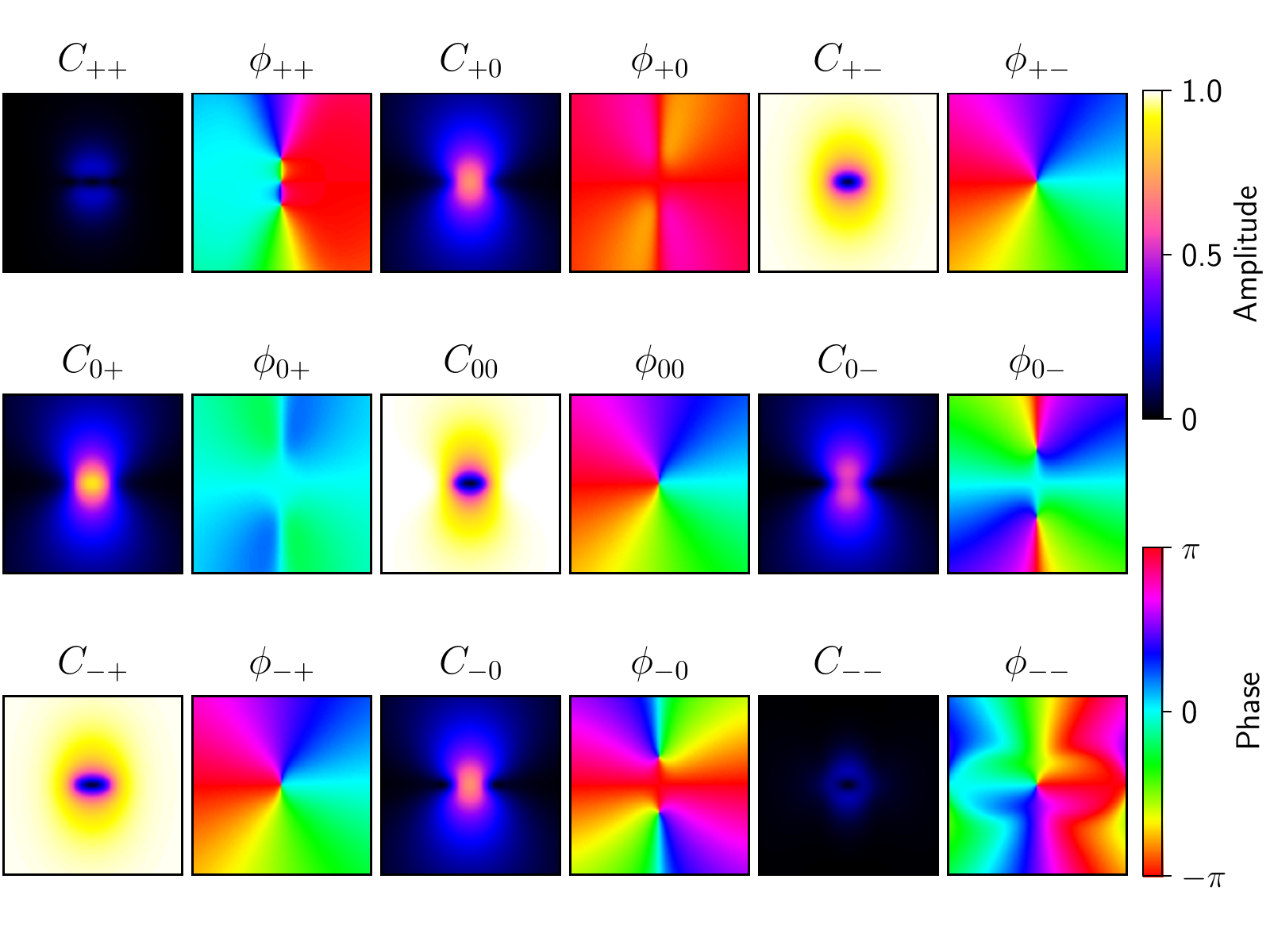}
\caption{
Amplitudes and phases of the components of the D-core vortex at $p=20$ bar and $T=0.55\,T_c$, shown and computed on the same grid as that in Fig.~\ref{fig-NCore_AmpPhase}. The key amplitudes and phases defining the D-core vortex are those with $N=2$ phase winding, $C_{0-}$ and $C_{-0}$. The D-core accommodates the double phase winding by dissociation into two $N=1$ vortices, allowing the corresponding amplitudes to grow. This is important for the stability of the D-core vortex relative to the A-core and o-vortex, as discussed in Sec.~\ref{sec-non-axial_symmetric_vortices}. The dissociation of the $N=2$ vortices is responsible for the broken axial symmetry that is clearly shown in all the amplitudes and phases. 
}
\label{fig-DCore_AmpPhase}
\end{figure}
%-------------------------------------------------

At sufficiently high pressure the strong-coupling corrections that stabilize the bulk A phase also stabilize the A-core vortex as the lowest energy vortex phase in \Heb. As a result the A-core vortex has a ``superfluid core'', with finite condensate density, $\Tr{AA^{\dag}}$ as shown in Fig.~\ref{fig-OP_Density}. Furthermore, the vortex circulation induces, via the Barnett effect,\cite{bar1915} a substantial spin polarization in the form of the $\beta$ phase, discussed in more detail in Sec.~\ref{sec-Intrinsic_Angular_Momentum}.

Figure~\ref{fig-ACore_AmpPhase} shows a stationary solution of Eqs.~\ref{eq-GL_equations} with axial symmetry which hosts both the chiral A-phase ($C_{0+}$) and $\beta$ phase ($C_{+0}$) with non-zero amplitudes in the vortex core. 
Note the large A-phase density, as well as the finite, but reduced, $\beta$-phase density in the core. Since there is no phase winding to suppress these amplitudes they grow to values near the corresponding homogeneous bulk values of a superposition of confined A and $\beta$ phases.
Thus, the ratio of the two condensate densities in the A-core vortex is of order $|C_{+0}(0)|^2/|C_{0+}(0)|^2\approx 0.1$ at $p =34$ bar based on the strong-coupling enhancement of the A phase as shown in Fig.~\ref{fig-ACore_AmpPhase}.

%----------------------------------------------------------------------------------------
\vspace*{-3mm}
\subsection{Non-Axial Symmetric Vortex States in \Heb}\label{sec-non-axial_symmetric_vortices}
\vspace*{-3mm}

Axial symmetry forces amplitudes with winding numbers $N=2$, i.e. $C_{0-}$ and $C_{-0}$, to be quadratically suppressed in the core as is shown in Fig.~\ref{fig-ACore_AmpPhase} for the A-core vortex (these amplitudes are zero by symmetry for the o-vortex).

For doubly quantized vortices the quadratic suppression of the core amplitude, combined with the cost in kinetic energy, generally leads to dissociation of doubly quantized vortices into a pair of singly quantized vortices in order to recover lost condensation energy for fixed total circulation.

Thus for the A-core vortex, if the amplitudes with $N=2$ winding numbers were to dissociate into a pair of $N=1$ vortices then the result would be a gain in condensation energy due to increased condensate amplitudes $C_{0-}$ and $C_{-0}$ in the core. 

%-----------------------------------------------------------------------------------------
\onecolumngrid
\begin{figure}[t]
\begin{minipage}{\textwidth}
\hspace*{-12mm}
\begin{tabular}{ccc}
\raisebox{-0.5\height}{\includegraphics[height=6.250cm]{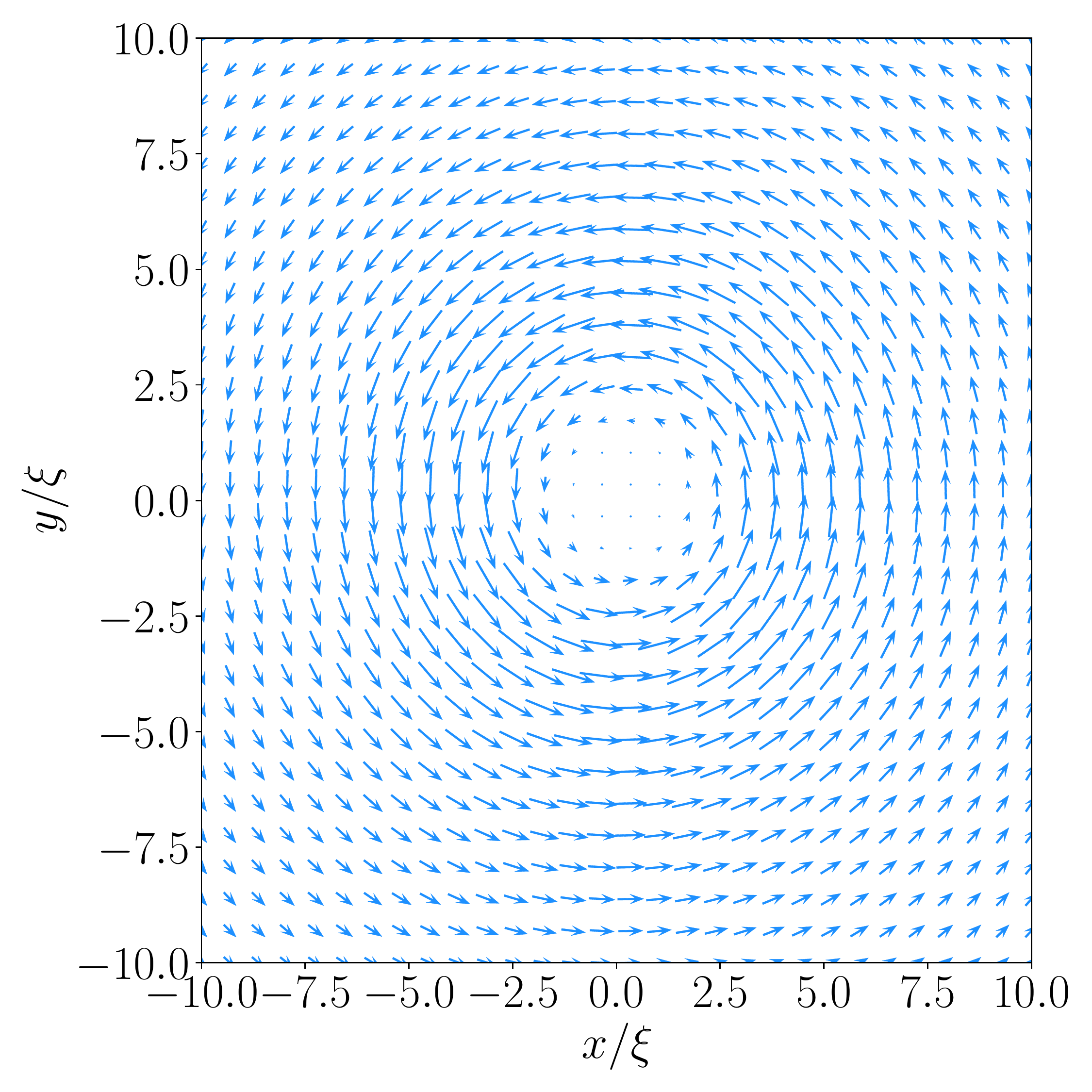}}
&
\hspace*{-5mm}
\raisebox{-0.5\height}{\includegraphics[height=6.250cm]{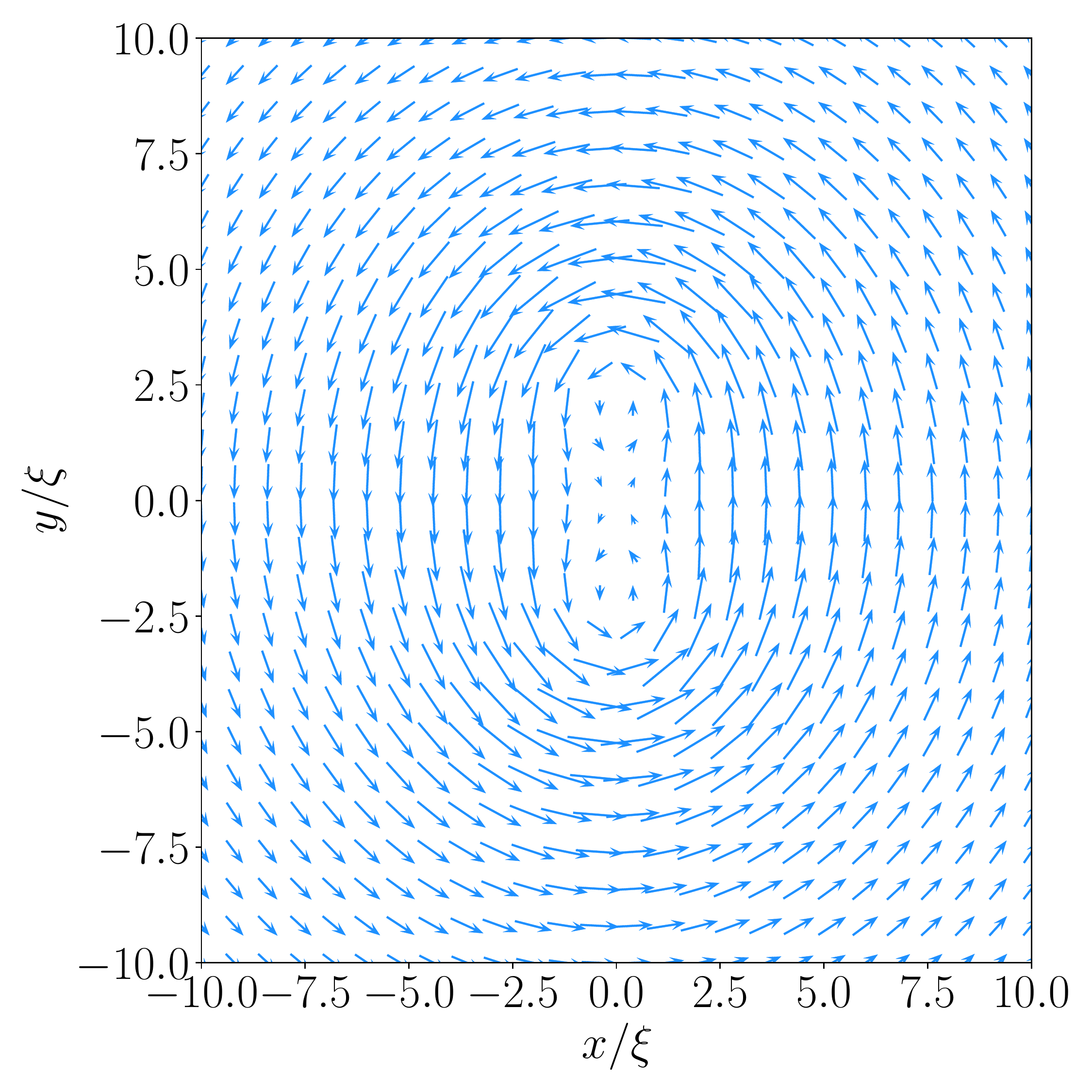}}
&
\hspace*{-5mm}
\raisebox{-0.5\height}{\includegraphics[height=6.250cm]{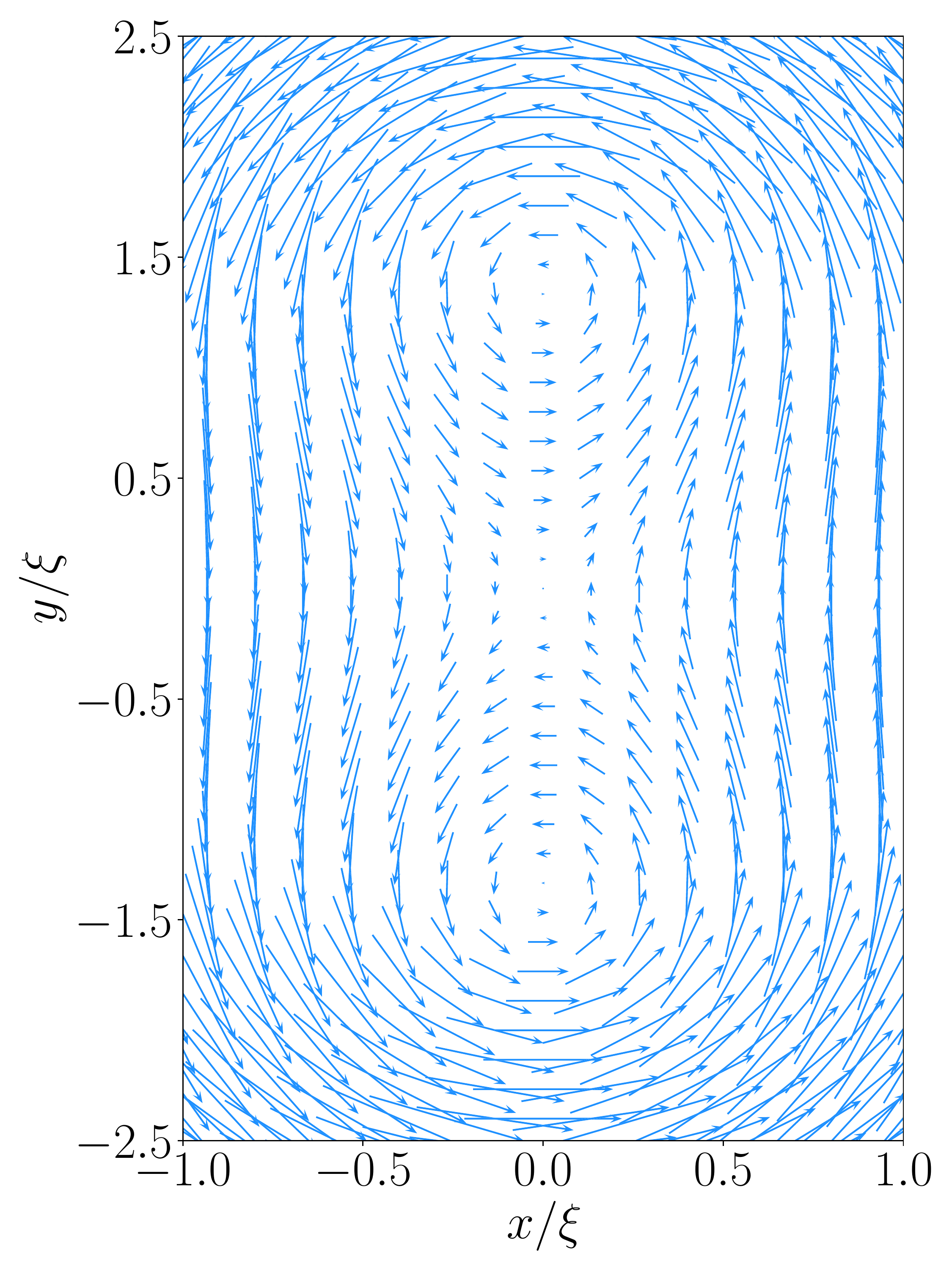}}
\end{tabular}
\caption{
Left: 
Axially symmetric current density of the A-core vortex for $p=34$ bar and $T=1.86$ mK.
The current is strongly suppressed to zero by the growth of the A- and $\beta$ 
phases in the core for $|\vr|\lesssim 2.5\xi$.
Center: 
Anisotropic mass current flow field of the D-core vortex at $p=20$bar, $T=1.23$ mK.
Right: 
Expanded view of the current near the center of the D-core showing the double 
vortex structure as the source of the anisotropic current density.
Currents are scaled in units of $j_c$ defined in Eq.~\ref{eq-critical_current}.
}
\label{fig-Vortex_Mass_Currents}
\end{minipage}
\end{figure}
\twocolumngrid
%-----------------------------------------------------------------------------------------

\noindent The cost of dissociation is the potential reduction in core energy from the amplitudes with zero phase winding. For the A-core vortex these amplitudes, $C_{0+}$ and $C_{+0}$, with $N=0$ are favorable because of strong-coupling energies. Thus, there is a competition between a gain in condensation energy by dissociation of the amplitudes with $N=2$ winding numbers and the loss in condensation energy of the $N=0$ amplitudes favored by strong-coupling and Zeeman energies. 
This competition is responsible for the stabilization of the D-core vortex as the temperature is lowered below $T_{\text{V}}(p,H)$, where strong-coupling energies are no longer sufficient to stabilize the axially symmetric A-core vortex, shown as the solid (dashed) green phase boundary for zero field ($H=284$ G) in Fig.~\ref{fig-Vortex_Phase_Diagram}.

At low pressures and low temperatures where strong coupling energies are relatively small the A-core vortex is no longer competitive with the D-core vortex. Furthermore, the o-vortex is never competitive with the D-core vortex, since forcing the $N=0,2$ components to vanish incurs too large a cost in condensation energy for the o-vortex compared to the D-core vortex, even in weak-coupling theory. 

The splitting of the $N=2$ vortices into a pair of $N=1$ vortices is shown clearly in the plots of the phases $\phi_{0-}$ and $\phi_{-0}$ in Fig.~\ref{fig-DCore_AmpPhase}, as is the growth in the amplitude for these components compared to their suppressed values in the A-core vortex. 
What is also clear is that the origin of the broken axial symmetry is the splitting of the $N=2$ phase singularities. This splitting of the $C_{0-}$ and $C_{-0}$ vortices along the $y$ axis breaks axial rotation symmetry, and generates a substantial uniaxial anistropy in the amplitudes $C_{0-}$ and $C_{-0}$, as well as all other components.
The connection between the broken axial symmetry of the D-core vortex and the dissociation of the $N=2$ vortices in $C_{0-}$ and $C_{-0}$ along the $y$ axis is particularly evident in the mass current distribution discussed below and shown in the right panel of Fig.~\ref{fig-Vortex_Mass_Currents}, where the pair of dissociated mass current vortices located at $y\approx \pm 1.5\,\xi$ dominate the internal structure of the D-core vortex mass current distribution.

\vspace*{-5mm} 
\subsection{Mass Current Density}\label{sec-Mass_Currents}
\vspace*{-3mm} 

Galilean invariance in pure \He\ has important implications for the transformation of velocities and mass currents in both normal and superfluid \He. In particular the order parameter transforms as $A_{\alpha i}(\vr)\xrightarrow[]{\vu}A_{\alpha i}(\vr)\,e^{-i2m\vu\cdot\vr/\hbar}$ under a Galilean boost with velocity $\vu$. Thus, the phase of the order parameter undergoes a local gauge transformation, or equivalently, $\vv_s\equiv(\hbar/2m)\mathbf{\grad}\vartheta$, transforms as a velocity field under a Galilean boost, $\vv_s\xrightarrow[]{\vu}\vv_s-\vu$. 
Galilean invariance also implies that the free energy density transforms as $f\xrightarrow[]{\vu}f-\vj\cdot\vu + \cO(u^2)$ where $\vj$ is the mass current density. For a boost from the rest frame of the normal excitations, i.e. $\vv_n = 0$, the gradient terms in the GL free energy density transform as $f_{\text{grad}}\xrightarrow[]{\vu}f_{\text{grad}}-\vj_s\cdot\vu+\mathcal{O}(u^2)$. Thus, by carrying out the boost transformation we obtain the superfluid mass current density in the rest frame of the excitations, expressed in terms of Cartesian components, 
\be
\hspace*{-2.65mm}
j_{s,i}\ns=\ns\frac{4m}{\hbar}\Im\ns
\left[
K_1\,A_{\alpha j}^{*}\grad_{\ns i} A_{\alpha j}
\ns+\ns
K_2\,A_{\alpha j}^{*}\grad_{\ns j} A_{\alpha i}
\ns+\ns
K_3\,A_{\alpha i}^{*}\grad_{\ns j} A_{\alpha j}
\right]
\ns.\ns
\ee

Far from the vortex core the phase gradient is small, $|\grad\vartheta| \ll \pi$, or equivalently the flow velocity is small compared to the maximum sustainable condensate velocity, i.e. $v_s\ll v_c=\hbar/\xi$. Thus, the current reduces to its value in the London limit governed by the local B phase order parameter in Eq. \ref{eq-p_quantized_B_vortex},
\be\label{eq-mass_current_London}
\vj_s
=
2\left(\frac{2m}{\hbar}\right)^2
\left(K_1+\nicefrac{1}{3}(K_2+K_3)\right)\,
\Delta_{\text{B}}^2\,
\vv_s
\,,
\quad v_s\ll v_c
\,.
\ee
The mass current recovers axial symmetry in the limit $|\vr| \rightarrow \infty$, however, the anisotropic corrections to axial flow decay slowly as $1/r^2$. Equation \ref{eq-mass_current_London} provides the characteristic scale for the vortex mass currents in rotating \Heb, 
\be\label{eq-critical_current}
j_c=2(2m/\hbar)^2(K_1+(K_2+K_3)/3)\Delta_{\text{B}}^2\,(\hbar/\xi)
\,.
\ee

Figure~\ref{fig-Vortex_Mass_Currents} shows the flow field for the mass current of both the A-core and D-core vortices. The A-core vortex has an axial vortex flow that collapses and vanishes rapidly in the zero phase-winding region of the A-phase and $\beta$-phase core, $|\vr|\lesssim 2.5\xi$, as shown in the left panel of Fig.~\ref{fig-Vortex_Mass_Currents}. By contrast the broken axial symmetry of the D-core vortex is evident in the center panel of Fig.~\ref{fig-Vortex_Mass_Currents}. A zoomed region of the anisotropic core is shown in the right panel which clearly shows the origin of the uniaxial anisotropy of the current flow is the dissociation of the $N=2$ vortices of $C_{0-}$ and $C_{-0}$ into a pair of $N=1$ vortices at $y\approx \pm 1.5\xi$. 

Another remarkable property of the D-core vortex, first reported in Ref.~\cite{thu87}, is the prediction of an axial current anomaly, i.e. the local pattern of currents flowing along the axis of circulation, but with zero net mass transport and zero phase gradient along the vortex axis~\footnote{A similar axial current anomaly bound to disclination lines in the chiral phase of \He\ confined in a cylindrical channel is reported in Ref.~\cite{wim18}}. The z-axis current density can be expressed as 

\begin{figure}[t]
\includegraphics[height=6.250cm]{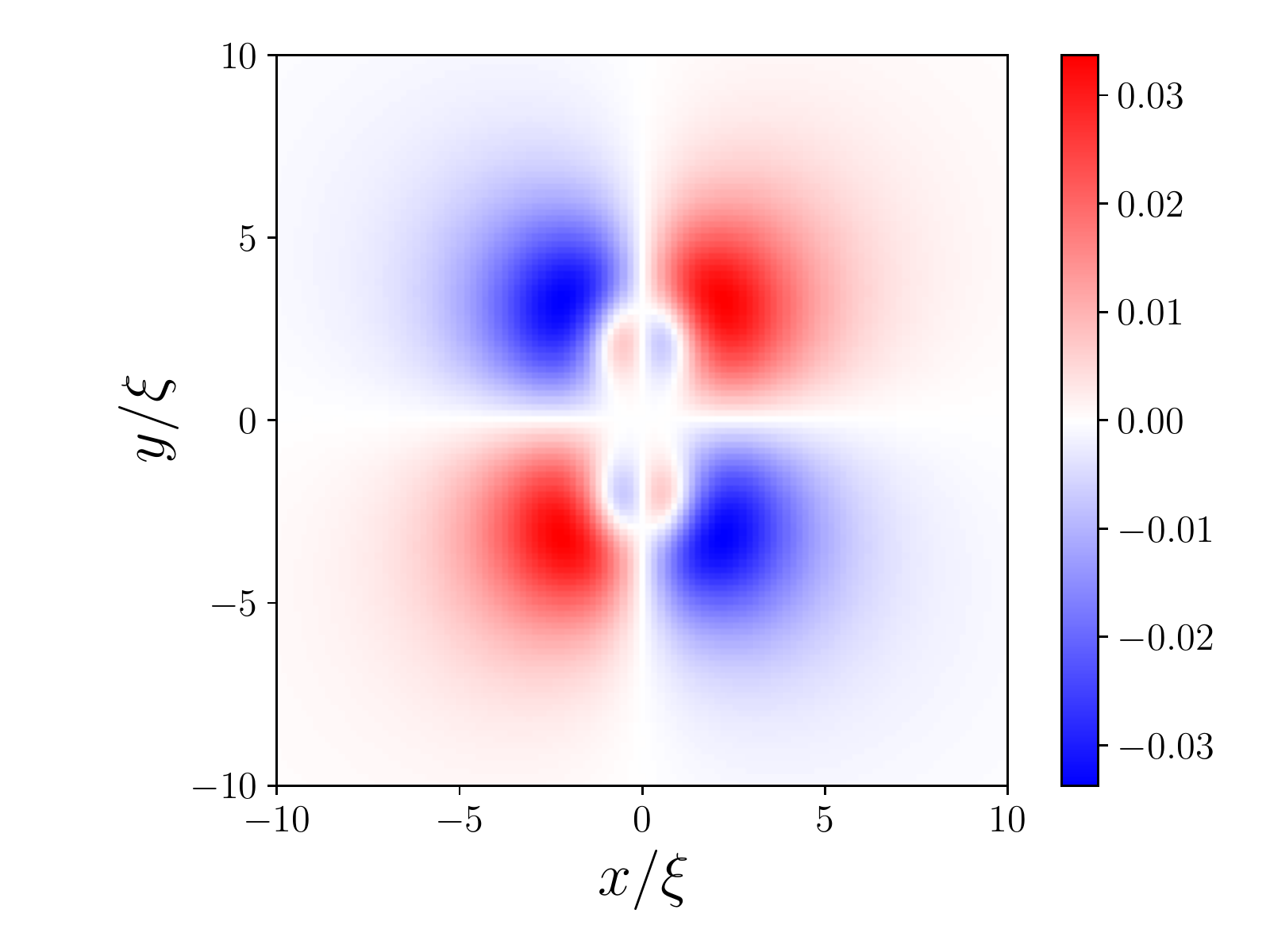}
\caption{
Axial mass current of the D-core vortex at $p=20$ bar and $T=0.55\,T_c$. The current density is
scaled in units of $j_c$.
}
\label{fig-Vortex_DC_Axial_Current}
\end{figure}

\ber
\hspace*{-5mm}
j_{s,z} 
\ns&=&\ns
\frac{4m}{\hbar}
\Im\ns
\left[
K_2\,A^*_{\alpha j}\grad_j A_{\alpha z}
\ns+\ns
K_3\,A^*_{\alpha z}\grad_j A_{\alpha j}
\right]
\\
\ns&=&\ns
\frac{4m}{3\hbar}\,\Delta_{\text{B}}^2\,
\Im\ns
\left[
K_2\,C^*_{\mu\nu}e^{i\nu \phi}\partial_\nu C_{\mu 0} 
\ns+\ns
K_3\,C^*_{\mu 0}\partial^*_\nu C_{\mu \nu }e^{-i\nu \phi}
\right]\ns,
\eer
where $\partial_0=\partial_z$ and $\partial_{\pm}=(\partial_x \mp i\partial_y)/\sqrt{2}$.
Figure~\ref{fig-Vortex_DC_Axial_Current} shows the z-axis current for the D-core vortex for the same pressure and temperature as that for the vortex currents in Fig.~\ref{fig-Vortex_Mass_Currents}, also in units of $j_c$. The axial current density spans an area of order $\cA \approx 100\,\xi^2\approx 6.25\,\mu\mbox{m}^2$.

An idea for detection of the current anomaly along the z-axis is to inject electrons into rotating \Heb\ from the outer, radial  boundary. Electrons in \Heb\ form mesoscopic ions of radius $R\approx 1.5\,\mbox{nm}$~\cite{she16}. The capture of these ions by D-core vortices should lead to transport of the ions along the D-core vortex lines driven by the axial currents. 
Detection of the ions by imaging on the top and bottom surfaces of the rotating vessel containing superfluid \Heb, in much the same way in which vortices in rotating superfluid \Hefour\ were first imaged~\cite{wil74}, would provide direct evidence of the axial mass currents~\footnote{Note that in contrast to the imaging of trapped ions in rotating \Hefour\ the transport of ions in the D-core phase of rotating \Heb\ would be under conditions of zero electric field.}.

%-----------------------------------------------------------------------------------------
%-----------------------------------------------------------------------------------------

\subsection{Intrinsic Vortex Orbital and Spin Angular Momentum}\label{sec-Intrinsic_Angular_Momentum}

To obtain the large scale structure of the vortex lattice for superfluid \Heb\ in equilibrium with a confining boundary rotating with a constant angular velocity, $\vOmega$, we must transform the free energy functional to the frame co-rotating with the boundary potential. Equilibrium in a rotating frame is achieved by a Legendre transformation~\cite{LL5}, $\cF'=\cF-\vJ\cdot\vOmega$, where $\vJ=\vL+\vS$ is the total angular momentum of liquid \He, including the orbital fluid angular momentum, $\vL$, and the nuclear spin angular momentum of \He. 
The Legendre transformation leads to several energy scales associated with equilibrium in a rotating frame. The dominant effect of $-\vJ\cdot\vOmega$ is the entrainment of the normal fluid into co-rotation with velocity $\vv_n=\vOmega\times\vr$. However, the superfluid velocity field is irrotational, and thus superfluid \Heb\ minimizes the kinetic energy and accommodates co-rotation by the formation of a lattice of quantized vortices. 
A key observation is that the orbital component of the Legendre transformation, $-\vL\cdot\vOmega$, is achieved by introducing a gauge potential, $\va=(2m/\hbar)(\vOmega\times\vr)$. In particular, the gradient energy transformed to the co-rotating frame can be written in terms of a kinetic energy density expressed in terms of the co-variant derivative of the order parameter, $\grad\rightarrow\cD=\grad-i\va$, and a coupling to the circulation of the gauge potential, $\grad\times\va=(4m/\hbar)\vOmega$, to the intrinsic orbital angular momentum of the Cooper pairs, $f_\mathrm{grad}'=f_\mathrm{kin}'+f_\mathrm{orbital}'$,
\ber
f_\mathrm{kin}'
&=& 
K_{1}
(\cD_k A_{\alpha j})^{*}(\cD_k A_{\alpha j})
\nonumber\\
&+& \nicefrac{1}{2}K_{s}
\left[ 
(\cD_j A_{\alpha j})^{*}(\cD_k A_{\alpha k})
+
(\cD_k A_{\alpha j})^{*}(\cD_j A_{\alpha k})
\right]\,,
\qquad
\label{eq-gradient_energy_rotating-frame}
\\
\vspace*{3mm}
f_\mathrm{orb}'
&\equiv& -\cL_{\text{orb}}\cdot\vOmega
= \frac{4m}{\hbar}\,K_{a}
\varepsilon_{ijk}\,
\Im\left(
A_{\alpha i}^* A_{\alpha j}
\right)
\vOmega_k
\,,
\label{eq-intrinsic_angular-momentum}
\eer
where $K_{s}=K_2+K_3$ and $K_{a}=K_2-K_3$. In weak-coupling theory $K_a^{\text{wc}} = 0$, however, particle-hole asymmetry and strong-coupling corrections give $K_a\approx(\kb T_c/E_f)^2\,K_1^{\text{wc}}$, and thus to an intrinsic orbital angular momentum of order $\cL_{\text{orb}}=\lambda_{\text{orb}}(n\hbar/4)\,(\Delta/E_f)^2$, with $\lambda_{\text{orb}}\sim \cO(\ln(E_f/\kb T_c))$. The intrinsic orbital angular momentum is too weak to affect the relative stability of the vortex phases at typical rotation speeds.

However, there are perturbations that are important in understanding the vortex structure and NMR signatures of the vortex phases of \Heb. In particular, in addition to the nuclear dipole energy there is a contribution to the nuclear Zeeman energy that is \emph{linear} in the external field defined by the invariant, $f'_z=-\vm\cdot\vH$, where in terms of Cartesian components,
\ber
m_{\delta} &=& g'_z\,\Im\left(AA^{\dag}\right)_{\alpha\beta}\,\epsilon_{\alpha\beta\delta}
\,,
\eer
where $\vm$ ($\vs\equiv \vm/\gamma$) is the intrinsic nuclear magnetization (spin) density of the Cooper pairs. The bulk B phase is time-reversal symmetric with $\vs_{\text{bulk}}\equiv 0$. Thus, the gyromagnetic effect observed in rotating \Heb\ is a manifestation of intrinsic spin polarization of vortices, driven by vortex currents in the core region. This is a vortex manifestation of the Barnett effect{~\cite{bar1915}}, discussed in the context of vortices in the $^3$P$_2$ neutron superfluid predicted to exist in the interiors of rotating neutron stars~{\cite{sau80,sau82a}}.

\begin{figure}[t]
\includegraphics[width=\columnwidth]{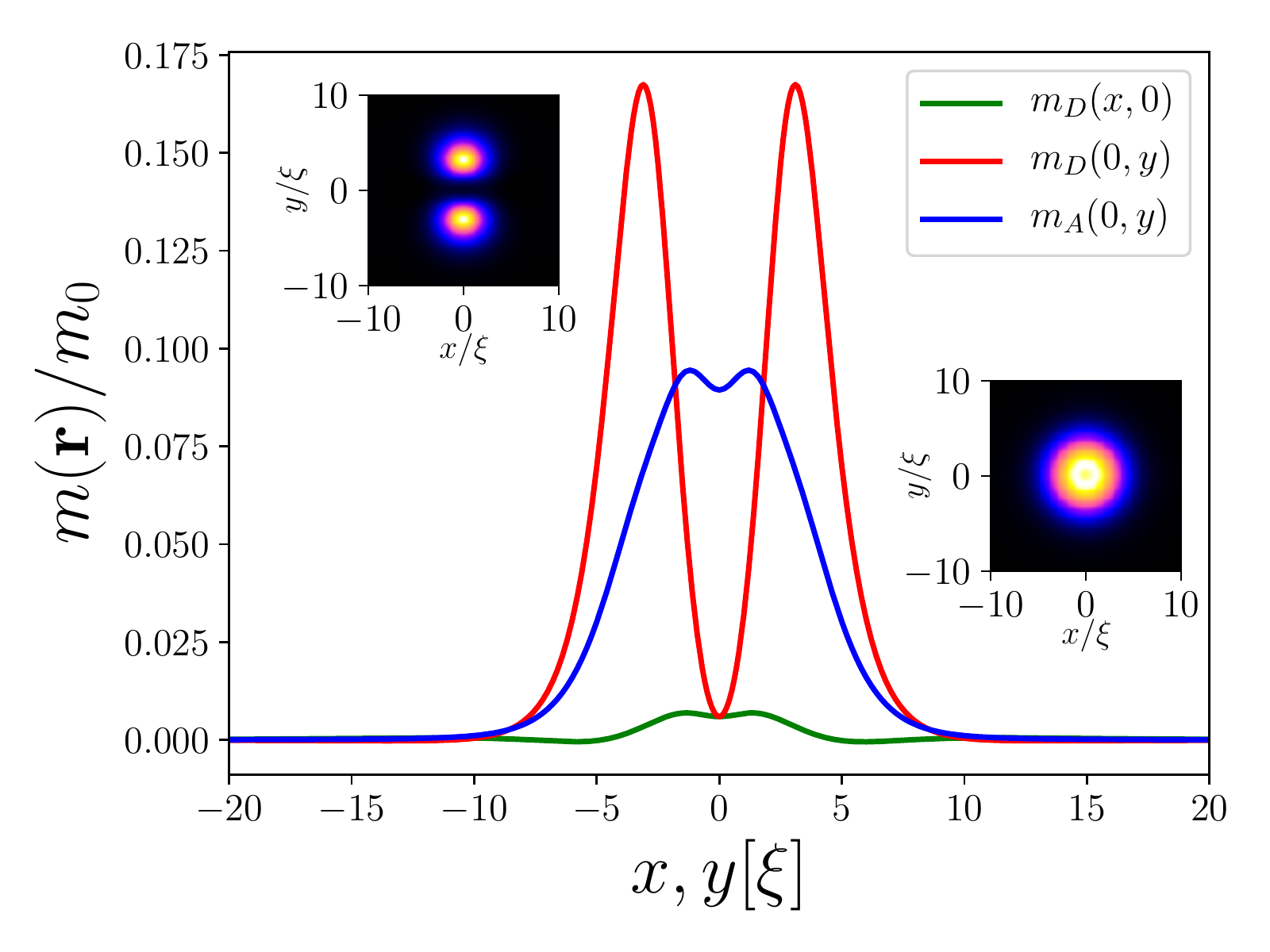}
\caption{
Magnetization profiles, $m_{A}$ and $m_{D}$, for the A- and D-core vortices, respectively. 
Insets: density plots of the same. 
For the A-core phase: $p=34$ bar $T=0.75\,T_c=1.86$ mK.
For the D-core phase: $p=20$ bar $T=0.55\,T_c=1.23$ mK.
}
\label{fig-Vortex_Magnetization}
\end{figure}

The intrinsic magnetization (spin polarization) for axially symmetric vortices takes a simple form when expressed in terms of amplitudes defined in the angular momentum basis,
\be
\vm(\vr)= m_0\sum_{\nu}\left(|C_{+\nu}|^2-|C_{-\nu}|^2\right)\,\hat\vOmega
+ 
\vm_{\perp}(\vr)
\,.
\ee
In addition to the axial component of the magnetization there is a transverse magnetization density, $\vm_{\perp}(\vr)$, which integrates to zero for all stationary vortex states, both the axially symmetric o-vortex and A-core vortex with $\vm_{\perp}=m_{\perp}(r)\hat\vr$, as well as the axially asymmetric D-core vortex. The magnitude of the intrinsic magnetization density is given by~\cite{sau80,sau82a},
\be
m_0\equiv g'_z\Delta_{\text{B}}^2\approx 
          n(\gamma\hbar)\ln(E_f/\kb T_c)(\Delta_{\text{B}}/E_f)^2
\,.
\ee
While all spin-triplet vortices generate an intrinsic spin polarization, symmetry constraints on the phase winding of the order parameter components that inhabit the vortex core, as well as strong-coupling terms in the free energy functional that stabilize vortex core states with zero phase winding, lead to vortex-core magnetic moments that reflect the symmetry of the vortex core order parameter. 
In the case of the high-pressure phases of rotating \Heb\ the A-core vortices, which host the ferromagnetic $\beta$ phase in the core, possess a substantial non-vanishing magnetization density in the cores. The D-core vortex phase also has a substantial vortex magnetization, which also reflects the double-core structure of that phase. The vortex magnetization density is shown for both phases in Fig.~\ref{fig-Vortex_Magnetization}. 
The total magnetic moment of the A- and D-core vortex phases, $\vM=\int d\vr\,\vm(\vr)$ exhibits a discontinuity at the first-order vortex phase transition. For example, the magnetization per unit length ($M/L_v$), per vortex jumps from $M_A/L_v=2.95\,m_0\xi^2$ in the A-core phase to $M_D/L_v = 5.36\,m_0\xi^2$ in the D-core phase at $T=2.0\,\mbox{mK}$ and $p=15.0\,\mbox{bar}$.

The direction of the vortex magnetization is selected by the angular velocity, i.e. $\vm_{\text{V}}=m(\vr)\hat\vOmega$. As a result the linear Zeeman energy, $f_z'=-m(\vr)\hat\vOmega\cdot\vH$ is the origin of the gyromagnetic effect observed in the NMR spectrum for the phases of rotating \Heb.

%-----------------------------------------------------------------------------------------
%-----------------------------------------------------------------------------------------

%-----------------------------------------------------------------------------------------
\begin{figure}[t]
\includegraphics[width=\columnwidth]{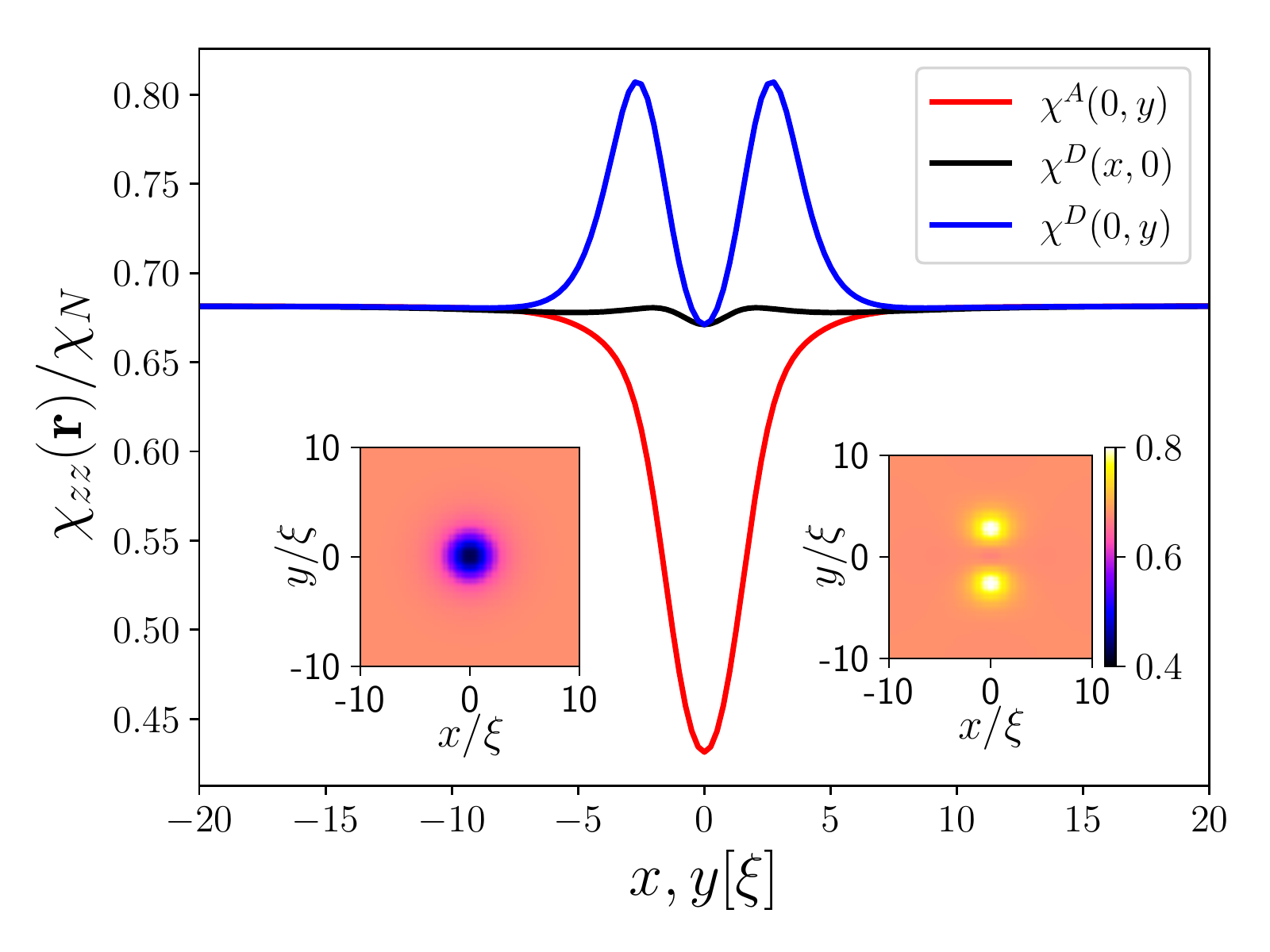}
\caption{
Susceptibility profiles, $\chi_{zz}^{A}$ and $\chi_{zz}^{D}$, for the A- and D-core vortices, respectively. Calculations are for the zero-field order parameters at $p=34\,\mbox{bar}$ just to the right and left of the zero-field transition line: $\chi_{zz}^A$ ($\chi_{zz}^D$) is evaluated at $T=1.80\,\mbox{mK}$ ($T=1.79\,\mbox{mK}$).
Insets: density plots of the same.
}
\label{AC+DC_Susceptibilities-zz}
\end{figure}
%-----------------------------------------------------------------------------------------

\section{Magnetic Susceptibility}\label{sec-Magnetic_Susceptibility_Vortex_States}

\begin{figure}[t]
\includegraphics[width=\columnwidth]{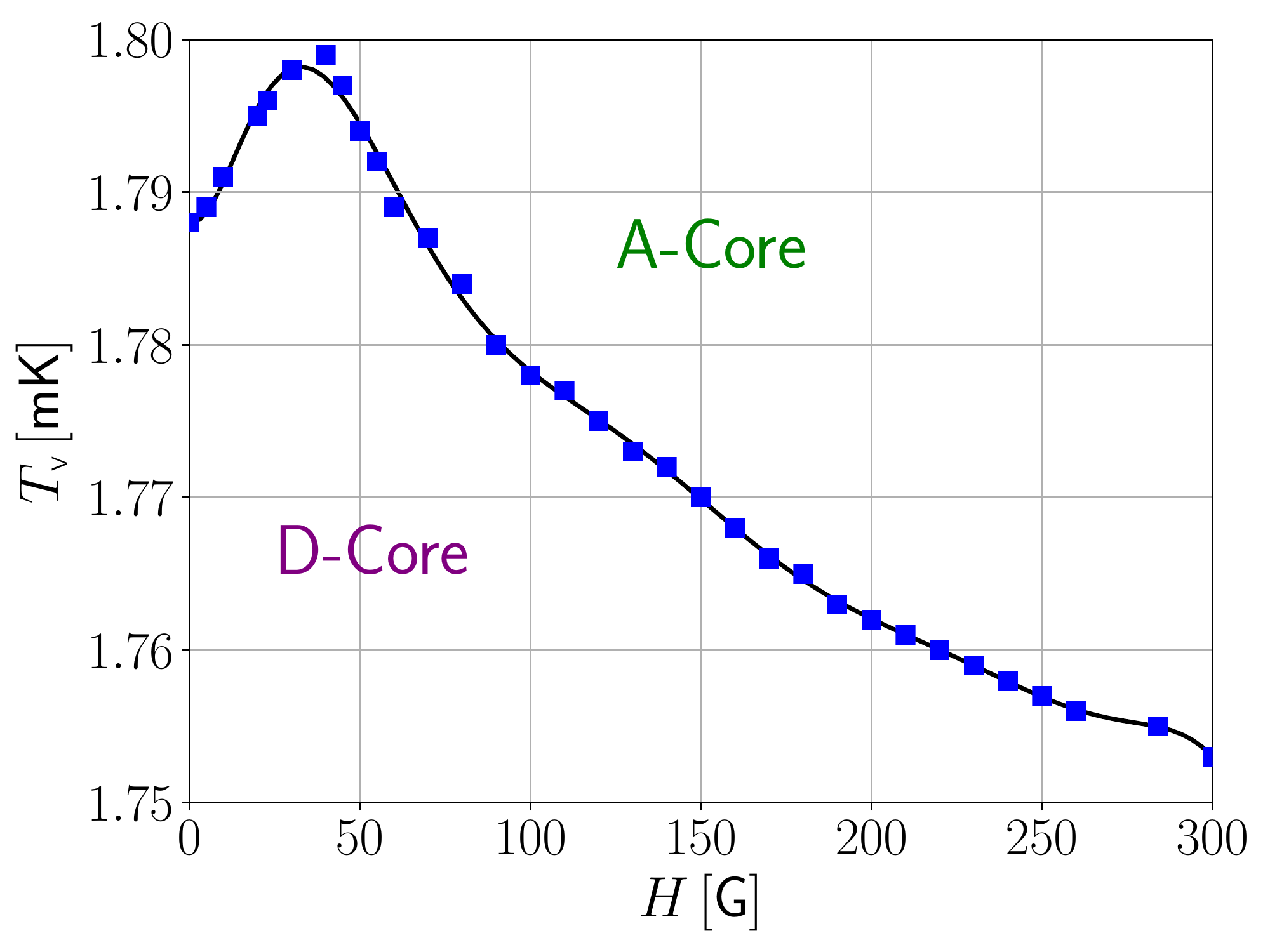}
\caption{
Field evolution of the equilibrium vortex transition temperature at $p=34\,\mbox{bar}$.
}
\label{fig-Vortex_Transition_vs_H}
\end{figure}

At sufficiently low magnetic fields the magnetization is determined by the nuclear Zeeman energy,
\be
\cF_{\text{Zeeman}} = -\nicefrac{1}{2}\int\,d^3r\,H_{\alpha}\chi_{\alpha\beta}(\vr)H_{\beta}
\,,
\ee
evaluated with the zero-field order parameter, i.e. neglecting order parameter distortion by the external field. For fields $\vH \parallel \vz$ the corresponding local magnetic susceptibility is
\be
\chi_{zz}(\vr)/\chi_{\text{N}} = 1 - 2g_z\,\sum_{i}\left|A_{zi}(\vr)\right|^2
\,.
\ee
Figure~\ref{AC+DC_Susceptibilities-zz} shows our results for the local susceptibilities of the A- and D-core vortices. The D-core vortex has the larger susceptibility, and thus we expect that the equilibrium vortex transition line to shift to higher temperatures with the application of a weak magetic field. 
This is indeed what we find from self-consistent solutions of the the GL equations when we include the Zeeman energy in Eq.~\ref{eq-Zeeman_energy}. Figure~\ref{fig-Vortex_Transition_vs_H} shows the evolution of the equilibrium vortex transition temperature with field, $T_{\text{V}}(p,H)$, for $p=34\,\mbox{bar}$. The initial increase of $T_{\text{V}}$ with field is indicative of susceptibilities for the A- and D-core vortices. However, at fields $H\gtrsim 40\,\mbox{G}$ the transition temperature reaches a maximum, then decreases with increasing field, such that $T_{\text{V}}(p,H=284\,\mbox{G}) = 1.755\,\mbox{mK}< T_{\text{V}}(p,H=0)=1.787\,\mbox{mK}$. The increase in $T_{\text{V}}$ relative to the zero-field transition for $H\gtrsim 60\,\mbox{mK}$ results from distortion on the vortex-core order parameters by the field, which dominates the Zeeman term even at relatively low fields due to the near degeneracy of the two vortex phases. This leads to the equilibrium vortex phase transition line, $T_{\text{V}}(p,H)$, for $H=284\,\mbox{G}$ shown in Fig.~\ref{fig-Vortex_Phase_Diagram}.
The equilibrium transition line, $T_{\text{V}}(p,H)$, as well as the supercooling transition line, $T_{\text{V}}^*(p,H)$, are reported for fields $\vH||\vOmega||\hat\vz$, and for the background B-phase order parameter given in Eq.~\ref{eq-boundary_condition_field}. Our results neglect the dipolar interaction within the computational cell, $\xi\ll d_c\ll\xi_{\text{D}}$, but include the effects of vortex counterflow and field-induced gap distortion.
The weak nuclear dipole energy of \Heb\ confined in the cylindrical experimental cell used in the rotating \Heb\ experiments reported in Ref.~\cite{hak83} introduces a large-scale texture of the background B-phase that varies on the scale of the cell radius, $R\approx 2.5\,\mbox{mm}$.\footnote{In particular, the anisotropy axis defining the B-phase order parameter in Eq.~\ref{eq-OP_B-phase_texture} aligns along $\vH||\vz$ in the center of the cell, but tilts away from the $z$-axis at an angle of $\beta(r)$, with $\beta(r)\simeq\beta_1\,r$, with $\beta_1\lesssim\nicefrac{\pi}{4}\,r/R$ for pressures $p\gtrsim 20\,\mbox{bar}$ and rotation speeds $\Omega\lesssim 2\,\mbox{rad}/s$. The texture leads to the transverse NMR shift and the spectrum of spin-wave bound states. The effects of vortex counter flow confined in the vortex cores, as well as field-induced gap distortion tune the slope, and thus are observable in the NMR spectrum~\cite{hak83}.}
A discussion of the effects of non-axial magnetic fields, as well the possibility of weak inhomogenous broadening from large scale textural effects, on the vortex phases and $p-T-H$ phase diagram is outside the scope of this article.

We note that Kasamatsu et al. recently published a report on the effects of non-axial magnetic fields on the structure of vortices in rotating \Heb~\cite{kas19}. They use the set of strong-coupling GL material parameters obtained from analysis of several experiments by Choi et al.~\cite{cho07}. However, their calculations are based on the standard GL free energy functional~\cite{thu87}. This limits their analysis of relative stability of vortex phases in \Heb\ to pressures \emph{below} the polycritical point pressure, $\pPCP=21.22\,\mbox{bar}$, and temperatures very close to $T_c$, thus precluding an analysis of the stability of phases over the experimentally relevant region of the $p-T-H$ phase diagram.
Our analysis, based on the strong-coupling GL theory discussed in Sec.~\ref{sec-GL_Theory}, allows us to explore the entire pressure range, and specifically the phase diagram \emph{above} the polycritical point pressure and temperatures below the bulk A-B transition, which is the region most relevant to the phases and phase transitions observed in rotating \Heb.

%-----------------------------------------------------------------------------------------
%-----------------------------------------------------------------------------------------

%----------------------------------------------------------------------------------------
\section{Equilibrium \& Metastability Transitions}\label{sec-Metastable_A-core_Phase}

The experimental transition between the two distinct vortex phases of rotating \Heb\ is hysteretic as shown in Fig. 1 of Ref.~\cite{pek84a}. The vortex phase transition on cooling occurs at much lower temperature than the phase transition on warming. This is indicated on the pressure-temperature phase diagram for $p=29.3\,\mbox{bar}$ the transition on cooling occurs at $T_{\text{V}}^{*}=1.43\,\mbox{mK}$ while the transition on warming occurs at a higher temperature which we estimate to be $T_{\text{V}}=1.81\,\mbox{mK}$. The latter was identified as the temperature at which the NMR satellite frequency splitting measured on warming merges with that measured on cooling. 
There is some uncertainty in this value because both A-core and D-core vortices are local minima of the free energy functional. Thus, on warming the heat flux of quasiparticles may heat the vortex cores and prematurely convert some D-core vortices to A-core vortices. Thus, a smooth extrapolation of the NMR splitting on warming yields $T_{\text{V}}\approx 1.85\,\mbox{mK}$, also indicated in Fig.~\ref{fig-Vortex_Phase_Diagram}.
This is the only data we found in the literature for the transition on warming.
The data for the transitions on cooling for all reported pressures was obtained from Fig. 2 of Ref.~\cite{pek84a}. All data reported in Fig.~\ref{fig-Vortex_Phase_Diagram} of this report was converted from the Helsinki temperature scale to the widely accepted Greywall scale according to $T_{\text{Greywall}} = 0.89\,T_{\text{Helsinki}}$~\cite{gre86}.
The transitions on cooling all exhibit a sharp drop in the NMR frequency at the same temperature independent of rotation speed. There is no further supercooling, indicating that $T_{\text{V}}^{*}$ is a global instability below which there is only one phase that is a local minimum of the free energy.

The theoretical results we report for the phase diagram in Fig.~\ref{fig-Vortex_Phase_Diagram} are based on precise numerical solutions of the strong-coupling GL equations for the vortex phases of rotating \Heb. 
The experimental transition on warming at $p=29.3$ bar and $H=284$ G is in close agreement with our determination of the equilibrium transition line, $T_{\text{V}}(p,H)$, at that pressure and field. We identify the warming transition as the equilibrium vortex phase transition, i.e. point in the $(p,T)$ plane where the free energies of the two phases are equal. 
This interpretation is based on our calculations of the free energies of the high-temperature, high-pressure A-core phase and the low-temperature, low-pressure D-core phase. 
In particular, the equilibrium transition line calculated as the locus of points where the A-core and D-core free energies are equal is shown in Fig.~\ref{fig-Vortex_Phase_Diagram} for zero field as the solid green line.
This transition line terminates on the bulk transition line at a triple point $[p_{v_c}$, $T_{c}(p_{v_c})]$. Thus, there is a window within the B phase where the A-core vortex phase is the equilibrium phase even in zero field, with the A phase and $\beta$ phase inhabiting the cores of vortices within the A-core phase. 
The A phase is able to grow within the B-phase vortex core because of the suppression of the B-phase amplitudes with winding number $N=1$: $C_{00}$, $C_{+-}$ and $C_{-+}$ and the absence of any suppression for the $N=0$ amplitudes: $C_{0+}$ and $C_{+0}$.
Thus, with strong-coupling support for the A-phase the A-core vortex is stabilized at sufficiently high pressure and high temperature in the region shaded in green in Fig~\ref{fig-Vortex_Phase_Diagram}.
Also shown is the equilbrium region of the A-core vortex phase for the field of $\vH=284\,\mbox{G}\,\hat\vz$. Note that the equilibrium region of the A-core phase is extended to lower temperatures (c.f. Fig.~\ref{fig-Vortex_Transition_vs_H}) and pressures within the range, $T_{\text{V}}(p,H)<T<T_{\text{AB}}(p,H)$, as shown by the dashed green line in Fig.~\ref{fig-Vortex_Phase_Diagram}.
Our analysis also shows that the region between the transition at $T_{\text{V}}(p,H)$ and the transition at $T_{\text{V}}^{*}(p,H)$ corresponds to the region in which the high-temperature A-core vortex phase is a metastable local minimum of the free energy, but is not the global minimum. Thus, the A-core phase \emph{supercools} to the lower temperature, $T_{\text{V}}^{*}$, below which the high-temperature A-core phase is globally unstable to the D-core phase.
Indeed, the observed transitions on cooling for $H=284$ G, over the pressure range $20\,\mbox{bar}\lesssim p\le 34\,\mbox{bar}$, agree well with our theoretically determined metastability transition, $T_{\text{V}}^*(p,H)$, at which the A-core vortex phase is globally unstable to the D-core vortex phase.

The supercooling transition at $T_{\text{V}}^{*}(p,H)$ shown as the dashed purple line in Fig.~\ref{fig-Vortex_Phase_Diagram}, and the much larger region of metastability of the A-core vortex phase (shaded in pink), was obtained by starting at high pressure and high temperature in the region of global stability of the A-core vortex phase, then lowering temperature slightly below $T_{\text{V}}(p,H)$, where the D-core vortex is the global miniumum,  and initializing the order parameter with the higher temperature A-core order parameter field plus a small admixture (``seed'') of the D-core order parameter, i.e. $A^{\text{init}}(p,T_{\text{new}})=A^{\text{A-core}}(p,T_{\text{last}}) + \epsilon\,A^{\text{D-core}}(p,T_{\text{new}})$, where $\epsilon\ll 1$. 
Throughout the region bounded by $T_{\text{V}}(p,H)$ and $T_{\text{V}}^*(p,H)$ (shown in pink) the vortex initialized with the D-core perturbation returned to the axially symmetric A-core phase. The supercooling transition, $T_{\text{V}}^*(p,H)$, was the locus of points where the A-core was globally unstable.
We note that results for the supercooling transition require a fine computational grid. For a coarse grid of $h=0.5\xi$ the supercooling transtion is lower than that shown in Fig.~\ref{fig-Vortex_Phase_Diagram}, but converges to the reported transition line for $h\lesssim 0.15\xi$. The phase transition lines shown in Fig.~\ref{fig-Vortex_Phase_Diagram} were obtained on a $60\xi\times 60\xi$ computational grid with grid spacing $h=0.1\xi$.
Our numerical annealing procedure used to identify the region of metastability of the A-core phase agrees remarkably well with the experimental results for the transition obtained on cooling, both in the magnitude of the supercooling at pressures above $p_{c_v}$, as well as the rapid cross over in slope of $T_{\text{V}}^*(p,H)$ with pressure at the lower pressures approaching $p_{c_v}$. However, our region of metastability does not extend as low in pressure as the experimentally reported transitions on cooling.
Our interpretation of the latter is that below $p_{c_v}$ strong-coupling energies are never able to stabilize the A phase in the vortex core, without assistance from the Zeeman energy. This results in the termination of the supercooling line on the equilibrium A-core vortex phase boundary at a pressure near $p_{c_v}$.
We are not able to resolve the origin of the discrepancy in the minimum pressure for the metastable A-core 
phase within the strong-coupling GL theory. 
Such a resolution may require new experiments under rotation with pressure sweeps, or perhaps implementation of the full quasiclassical strong-coupling free energy functional extended to inhomogeneous phases.

%----------------------------------------------------------------------------------------
\vspace*{-3mm}
\section{Summary and Outlook}
\vspace*{-3mm}

The recent development of a strong-coupling Ginzburg-Landau theory that accounts for the relative stability of the bulk A and B phases has provided the first opportunity to examine the relative stability of the vortex phases discovered in rotating \Heb\ and to predict, based on known material properties of superfluid \He\ over the full pressure range, the equilibrium and metastable vortex phase transitions. We are able to verify the local and global stability of all the stationary solutions to the strong-coupling GL theory over the full $(p,T)$ plane. Only the A-core and D-core phases are global minima anywhere in the $(p,T)$ plane.
The results we report provide strong theoretical support for the identification of the experimentally observed phase transitions as the equilibrium and supercooled phase transitions between the high temperature A-core vortex phase with broken time-reversal and mirror symmetries (proposed by Salomaa and Volovik~\cite{sal83}), and the low temperature, low pressure D-core vortex phase with broken axial symmetry (proposed by Thuneberg~\cite{thu86}). Furthermore, both of these transitions are driven by the decrease in strong coupling energies at sufficiently low pressures and temperatures defined by the metastability line $T_{\text{V}}^{*}(p,H)$. In addition, the broken rotational symmetry of the D-core vortex is identified with the instability of the components within the core with $4\pi$ phase winding. Once strong-coupling energies are suppressed by sufficiently low temperature or pressure the doubly quanitized vortices dissociate to gain condensation energy, and as a result break axial symmetry.  

We conclude with the two forward looking observations.
First, the success of the strong coupling GL theory, evident by the results for the vortex phase diagram, provides a theoretical tool for studying a wide range of problems involving inhomogeneous phases with complex symmetry breaking and/or novel topological defects, in the strong-coupling limit, that were not previously accessible. A recent example is the analysis of the experimentally measured Bosonic collective mode frequencies (``Higgs masses'') of superfluid \Heb\ using a time-dependent extension of the strong-coupling GL theory in Ref.~\cite{sau17}, which provided consistent experimental results for the strength of the f-wave pairing interaction in superfluid \He\ over the full pressure range~\cite{ngu19}, a material parameter that is important for understanding ground-states and excitations of superfluid \He\ at high pressures and high magnetic fields.
Secondly, the strong-coupling GL theory is supported by the microscopic strong-coupling pairing theory based on leading order corrections to the weak-coupling BCS theory originating from binary collision scattering between fermionic quasiparticles of the normal phase of liquid \He~\cite{wim19}. Further development of a quantitative microscopic strong-coupling pairing theory to inhomogeneous, non-equilibrium states is well within reach.

%----------------------------------------------------------------------------------------
\vspace*{-3mm}
\section{Acknowledgements} 
\vspace*{-3mm}
A preliminary report of these results was presented at the International Conference on Quantum
Fluids and Solids (QFS2018), Tokyo, Japan in July 2018.
We thank Wei-Ting Lin for discussions on efficient numerical approaches to solving multi-component Euler-Lagrange PDEs, and Erkki Thuneberg for detailed comments on a preliminary version of this manuscript. 
The research was supported by the National Science Foundation (Grant DMR-1508730).

%----------------------------------------------------------------------------------------
\section{Appendix: Material Parameters}\label{app-material_parameters}

The following tables summarize the pressure dependent material parameters that determine the 
properties of the superfluid phases in strong-coupling theory.
 
%----------------- Wiman-Sauls Beta Parameters ---------------------------------------
\begin{table}[h]
\begin{center}\tiny
\begin{tabular}{|l|l|l|l|l|l|}
\hline
$n$ & \qquad \ $\beta^{sc}_1$ & $\qquad \ \beta^{sc}_2$ & \qquad \ $\beta^{sc}_3$ & \qquad \ $\beta^{sc}_4$ & \qquad \ $\beta^{sc}_5$ \\ \hline
    0 & $-9.849\times 10^{-3}$&  $-4.193\times 10^{-2}$  & $-1.322 \times 10^{-2}$ & $-4.747\times 10^{-3}$ & $-8.987\times 10^{-2}$ \\
    1 & $-5.043\times 10^{-2}$ & $-1.177\times 10^{-1}$ & $-5.428\times 10^{-2}$ & $-3.788\times 10^{-1}$ & $-6.925\times 10^{-1}$ \\
    2 & $ \ \ 2.205\times 10^{-2}$ & $-4.322\times10^{-2}$ & $\ \ 9.559\times 10^{-2}$ & $-1.774\times 10^{-1}$ & $\ \ 8.761\times 10^{-1}$ \\
    3 & $-2.557\times 10^{-2}$ & $ \ \ 8.793\times10^{-2}$ & $-6.419\times 10^{-2}$ & $\ \ 1.735\times 10^{-1}$ & $-5.929\times 10^{-1}$ \\
    4 & $\ \ 5.023\times 10^{-2}$ & $-8.598\times 10^{-2}$ & $- 9.310\times 10^{-3}$ & $\ \ 1.878\times 10^{-1}$ & $\ \ 2.904\times 10^{-2} $\\
    5 & $-2.769\times 10^{-2}$ & $\ \ 3.639\times 10^{-2}$ & $\ \ 1.862\times 10^{-2}$ & $-1.522\times 10^{-1}$ & $\ \ 8.870\times 10^{-2} $ \\ \hline
\end{tabular}
\caption{Coefficients of a polynomial fit to the strong-coupling $\beta$ parameters from 
Ref.~\cite{wim19} of the form $\scb{i} = \sum_n a^{(i)}_n \, p^n$.}
\label{table-strong-coupling-betafit}
\end{center}
\end{table}

%--------------------------- Polynomial fit to the Wiman-Sauls Betas ------------------
\begin{table}[h]
\begin{center}\tiny
\begin{tabular}{|c|c|c|c|c|c|c|c|c|c|c|c|}
\hline
$p$[bar] & $n$ [nm$^{-3}$] & $m^*/m$ & $F_0^a $ & $T_c$[mK] & $v_f$[m/s] & $\xi_0$[nm] & $\beta^{sc}_1$  & $\beta^{sc}_2$  & $\beta^{sc}_3 $  &$\beta^{sc}_4$ & $\beta^{sc}_5$ \cr
\hline
    0.0   & 16.28 & 2.80 & -0.7226 & 0.929 &  59.03 & 77.21 & -0.0098 & -0.0419 &  -0.0132 & -0.0047 & -0.0899\cr
    2.0   & 17.41 & 3.05 & -0.7317 & 1.181 &  55.41 & 57.04 & -0.0127 & -0.0490 & -0.0161 & -0.0276 & -0.1277\cr
    4.0   & 18.21 & 3.27 & -0.7392 & 1.388 &  52.36 & 45.85 & -0.0155 & -0.0562 & -0.0184 & -0.0514 & -0.1602\cr
    6.0   & 18.85 & 3.48 & -0.7453 & 1.560 &  49.77 & 38.77 & -0.0181 & -0.0636 & -0.0202 & -0.0760 & -0.1880\cr
    8.0   & 19.34 & 3.68 & -0.7503 & 1.705 &  47.56 & 33.91 & -0.0207 & -0.0711 & -0.0216 & -0.1010 & -0.2119\cr
   10.0  & 19.75 & 3.86 & -0.7544 & 1.828 &  45.66 & 30.37 & -0.0231 & -0.0786 & -0.0226 & -0.1260 & -0.2324\cr
   12.0  & 20.16 & 4.03 & -0.7580 & 1.934 &  44.00 & 27.66 & -0.0254 & -0.0861 & -0.0233 & -0.1508 & -0.2503\cr
   14.0  & 20.60 & 4.20 & -0.7610 & 2.026 &  42.51 & 25.51 & -0.0275 & -0.0936 & -0.0239 & -0.1751 & -0.2660\cr
   16.0  & 21.01 & 4.37 & -0.7637 & 2.106 &  41.17 & 23.76 & -0.0295 & -0.1011 & -0.0243 & -0.1985 & -0.2801\cr
   18.0  & 21.44 & 4.53 & -0.7661 & 2.177 &  39.92 & 22.29 & -0.0314 & -0.1086 & -0.0247 & -0.2208 & -0.2930\cr
   20.0  & 21.79 & 4.70 & -0.7684 & 2.239 &  38.74 & 21.03 & -0.0330 & -0.1160 & -0.0249 & -0.2419 & -0.3051\cr
   22.0  & 22.96 & 4.86 & -0.7705 & 2.293 &  37.61 & 19.94 & -0.0345 & -0.1233 & -0.0252 & -0.2614 & -0.3167\cr
   24.0  & 22.36 & 5.02 & -0.7725 & 2.339 &  36.53 & 18.99 & -0.0358 & -0.1306 & -0.0255 & -0.2795 & -0.3280\cr
   26.0  & 22.54 & 5.18 & -0.7743 & 2.378 &  35.50 & 18.15 & -0.0370 & -0.1378 & -0.0258 & -0.2961 & -0.3392\cr
   28.0  & 22.71 & 5.34 & -0.7758 & 2.411 &  34.53 & 17.41 & -0.0381 & -0.1448 & -0.0262 & -0.3114 & -0.3502\cr
   30.0  & 22.90 & 5.50 & -0.7769 & 2.438 &  33.63 & 16.77 & -0.0391 & -0.1517 & -0.0265 & -0.3255 & -0.3611\cr
   32.0  & 23.22 & 5.66 & -0.7775 & 2.463 &  32.85 & 16.22 & -0.0402 & -0.1583 & -0.0267 & -0.3388 & -0.3717\cr
   34.0  & 23.87 & 5.82 & -0.7775 & 2.486 &  32.23 & 15.76 & -0.0413 & -0.1645 & -0.0268 & -0.3518 & -0.3815\cr
\hline
\end{tabular}
\end{center}
\caption{Material parameters for \He{} vs. pressure, with 
the particle density $n=k_f^3/3\pi^2$ from Ref. \cite{whe75},
the effective mass, $m^{*}$, and $T_c$ from Ref.~\cite{gre86},
the exchange interaction, $F_0^{a}$, is from Ref.~\cite{gou06},
the Fermi velocity, $v_f=\hbar k_f/m^*$, calculated from the Fermi wavelength, $k_f$,
and 
the coherence length is $\xi_0=\hbar v_f/2\pi\,k_B T_c$.
The strong-coupling parameters, $\beta^{\text{sc}}_i$,
in units of $|\beta_1^{\text{wc}}|$, are from Ref.~\cite{wim19}.
}
\label{table-material_parameters}
\end{table}

%----------------------------------------------------------------------------------------
\section{Appendix: Numerical Methods}\label{app-l-bfgs}

%---------------------------------------------------------------------------------
\begin{figure}[t]
\centering
\includegraphics[width=\columnwidth]{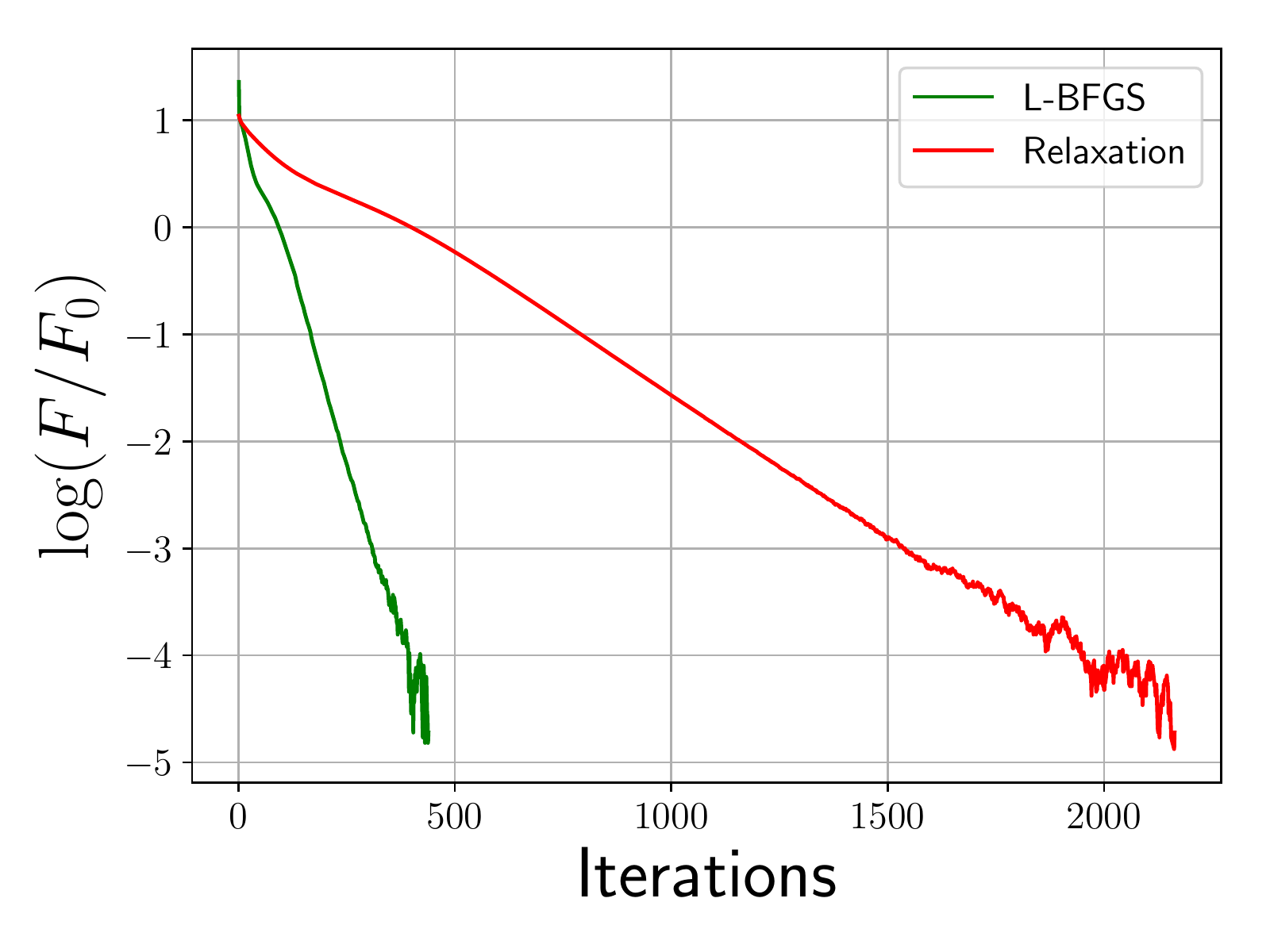}
\caption{Rate of convergence of the iterative solution of the GL equations expressed in terms of the 
Free energy at each interation, $k$, for the o-vortex at $p=10$ bar, $T=0.25\,T_c$ normalized by the bulk B phase free energy integrated over the same volume. The L-BFGS algorithm is shown in green, while the relaxation algorithm is shown in red.
}
\label{fig-Convergence}
\end{figure}
%---------------------------------------------------------------------------------

The search for stationary states of the Ginzburg-Landau functional leads to the Euler-Lagrange Equations (Eqs~\ref{eq-GL_equations}), which are coupled, non-linear partial differential equations (PDEs) for the 18 components of the \He\ order parameter.
The method of \emph{relaxation} based on the discretized version of Eq.~\ref{eq-TDGL_equation} to improve the approximate solution at each step along the gradient direction until one reaches the steady state solution is generally inefficient. 
Instead, we employ an efficient numerical method developed to solve the multi-component field equations, e.g. the order parameter for topological defects in superfluid \He.
The method is based on the L-BFGS optimization algorithm~\cite{noc18} summarized below.

In Newton's method, we solve the equation
\be\label{eq-Newton}
x_{k+1}=x_k + \alpha_k p_k
\ee
where $\alpha_k\equiv \alpha$ represents a fixed step size at each step labeled by $k$, and $p_k = -H^{-1}_kG_k$ is the search direction where $G_k$ is the functional gradient defined in Eq.~\ref{eq-GL_equations}. Hereafter we follow standard notation~\cite{noc18} and denote the inverse Hessian simply by $H_{k}$.
We implement this algorithm by storing the order parameter, $A_{\alpha i}(x,y)$ as a four-dimensional ($\alpha$,$i$,$x$,$y$) array of complex numbers represented by $x_k$ at step $k$. At each iteration, the order parameter is updated along with the step size and search direction. Storing the exact inverse Hessian, $H_k$, requires calculating a matrix of $N\times N$ second derivatives which is computationally expensive. Instead, we use the L-BFGS quasi-Newton minimization algorithm. This requires us to solve Eq.~\ref{eq-Newton} to determine a step size, $\alpha_k$, that minimizes the function $f(\alpha_k\hat{p}+x_k)$ at each iteration, where the direction $\hat{p}$ is constructed from the approximation to the inverse Hessian $H_k$, using $G_k$ and $x_k$. We require the inverse Hessian to be symmetric and positive definite. To determine $H_{k+1}$ we solve the minimization problem
\begin{eqnarray}
\min_H || H_k-H||,\quad H=H^\dagger, H y_k = s_k,
\end{eqnarray}
where $s_k = x_{k+1}-x_k, y_k = G_{k+1}-G_k$, where $G_k$ is the functional gradient of the GL functional at iterate $k$, $G_k=\delta F_k/\delta A^*_{\alpha i,k}$.
The unique solution to this minimization problem is obtained by re-writing the minimization problem in terms of a weighted Frobenius norm, which transforms the minimization problem to a new basis under a unitary transformation. The Frobenius norm can then be calculated explicitly and minimized. Transforming back to the original basis we obtain the  solution
\be\label{eq-BFGS}
H_{k+1}=(1-\rho_{k}s_{k}\,y^{\dag}_{k}) H_{k} (1-\rho_{k}y_{k}s^{\dag}_{k})+\rho_ks_k s^\dag_k
\,,
\ee

\vspace*{3mm}

\noindent where $\rho_k=1/(y^\dagger_k s_k)$. Eq.~\ref{eq-BFGS} is known as the BFGS update~\cite{noc18}, and is an approximation to the inverse Hessian $H_{k+1}$ given an initial inverse Hessian $H_k$.
Thus, we now solve Eq.~\ref{eq-Newton} with a search direction given by $p_k=-H_{k}G_k$ which is calculated in terms of inner products of the form $\langle y_k|G_k\rangle, \langle s_k|G_k\rangle$. 
This makes the solution of Eq.~\ref{eq-GL_equations} straight-forward with
\be
s_k=x_{k+1}-x_k,\qquad y_k=G_{k+1}-G_k
\,.
\ee

We initialize the inverse Hessian with $H_0 =1$, then update according to
\be
H_k=\frac{s^\dagger_{k-1}y_{k-1}}{y^\dagger_{k-1}y_{k-1}}.
\ee
This is an approximation to the inverse Hessian matrix along the most recent search direction. For the L-BFGS update at iteration $k$ we have the current iterate as $x_k$ and we store a limited memory set of vector pairs $\{s_i,y_i\}$ for $i=k-m,..,k-1$. Thus, by choosing an initial approximate inverse Hessian $H^0_k$ we obtain by repeated iteration of Eq.~\ref{eq-BFGS} the L-BFGS algorithm~\cite{noc18},
\ber
H_k
&\ns=\ns&
(V^\dagger_{k-1}\cdots V^\dagger_{k-m})\,H^0_k\,(V_{k-m}\cdots V_{k-1})
\\
&\ns+\ns&
\rho_{k-m}(V^\dagger_{k-1}\cdots V^\dagger_{k-m+1})s_{k-m}s^\dagger_{k-m}(V_{k-m+1}\cdots V_{k-1})
\nonumber\\
&\ns+\ns&
\rho_{k-m+1}(V^\dagger_{k-1}\cdots V^\dagger_{k-m+2})s_{k-m+1}s^\dagger_{k-m+1}(V_{k-m+2}\cdots V_{k-1})
\nonumber\\
&\ns+\ns&
\ldots +
\rho_{k-1}\,s_{k-1}\,s^{\dag}_{k-1}
\,.
\nonumber
\eer

The arrays $s_k, y_k$ which encode the order parameter and functional gradient are stored as five-dimensional complex arrays where one component of the array is a memory index and the other four components represent the orbital, spin and spatial degrees of freedom in the $x-y$ plane.
The L-BFGS algorithm is used to calculate the stationary states by solving Eq.~\ref{eq-GL_equations} for the full $(p,T)$ plane. In Fig.~\ref{fig-Convergence} we compare numerical relaxation with the rate of convergence of the L-BFGS algorithm for the axially symmetric o-vortex.
The performance of the L-BFGS algorithm is essential in being able to calculate the equilibrium and metastable phase diagram on reasonable timescales.

\section{Appendix: Benchmarking the GL Solver}\label{app-benchmark-GL}

\begin{figure}[h]
\centering
\includegraphics[width=\columnwidth]{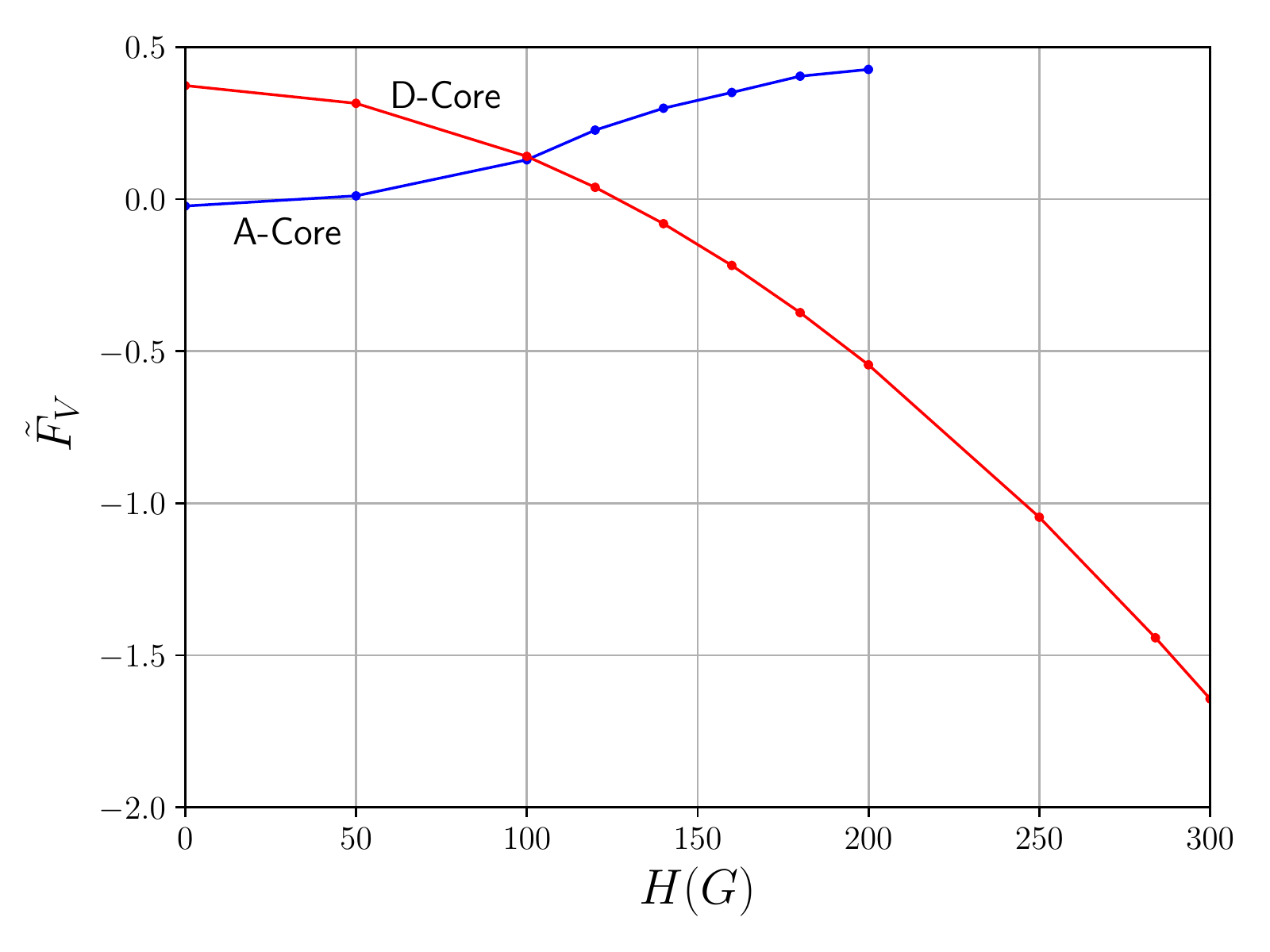}
\caption{
Free energies for the A-core and D-core vortex states versus magnetic field calculated using our GL solver for the set I GL parameters at $p=34\,\mbox{bar}$.
The core contribution to the free energies are calculated as in Ref.~\onlinecite{kas19} by subtracting the bulk and hydrodynamic contributions to the B phase energy, and normalizing the result in units of the bulk B phase energy at each field.
}
\label{fig-benchmark}
\end{figure}

We tested our code against others by comparing our results for the free energies of the A-core and D-core vortex states with those reported in Ref.~\onlinecite{kas19} based on their choice for the GL $\beta$ parameters. Figure~\ref{fig-benchmark} shows results based on our GL solver using GL parameter set I of Ref.~\onlinecite{kas19} for $p=34\,\mbox{bar}$. Our result is in excellent agreement with the result reported by Kasamatsu et al. in their Fig. 5(d) for the same GL parameters, including the crossing field of $H\approx 100\,\mbox{G}$. N.B. While these are local minima of the GL functional for this parameter set, they do not represent realized solutions at this pressure because this set of GL $\beta$ parameters does not account for the relative stability of the bulk A and B phases for pressures above $\pPCP=21.22\,\mbox{bar}$. Nevertheless, the comparison provides a benchmark and additional confidence in our GL solver and numerical results.

%\newpage

%---------------------------------------------------------------------------------
%---------------------------------------------------------------------------------

%---------------------------------------------------------------------------------
%\bibliography{QFS,Books,ASTRO,CM,Numerics}
%---------------------------------------------------------------------------------
%
\end{document}